\documentclass[twocolumn]{aastex631}

\received{November 24, 2021}
\accepted{December 22, 2021}

\submitjournal{AJ}

\shorttitle{MeerKAT Primary Beam Measurements}
\shortauthors{de Villiers \& Cotton}

\usepackage{amsmath}
\usepackage{multirow}
\graphicspath{{./}{figures/}}

\begin{document}

\title{MeerKAT Primary Beam Measurements in \em L\em -band}

\correspondingauthor{Mattieu S. de Villiers}
\email{mattieu@sarao.ac.za}

\author[0000-0001-5628-0417]{Mattieu S. de Villiers}
\affiliation{South African Radio Astronomy Observatory,
2 Fir Street, Black River Park, Observatory, 7925, RSA}

\author[0000-0001-7363-6489]{William D. Cotton}
\affiliation{National Radio Astronomy Observatory, 520 Edgemont Road, Charlottesville, VA 22903, USA}

\begin{abstract}

Full-polarization primary beam patterns of MeerKAT antennas have been measured in \em L-\em band (856 to 1711 MHz) by means of radio holography using celestial targets. This paper presents the observed frequency dependent properties of these beams, and guides users of this 64 antenna radio telescope that are concerned by its direction dependent polarization effects. In this work the effects on the primary beams due to modeling simplifications, bandwidth averaging, gravitational loading and ambient temperature are quantified within the half power region of the beam. A perspective is provided on the level of significance of typical use case effects. It is shown that antenna pointing is a leading cause of inaccuracy for telescope users in the presumed beam shape, introducing errors exceeding 1\% in power near the half power point of beams, owing to a telescope pointing accuracy of $\sigma\approx 0.6$ arcminutes. Disregarding these pointing errors, variability in the Stokes I beam shape relative to the array average is most commonly around 0.3\% in power; however, the impact above 1500 MHz is on average triple that of the lower half of the band. This happens because the proportion of higher order  waveguide modes that are activated and propagate is sensitive to small manufacturing differences in the orthomode transducer for each receiver. Primary beam correction verification test results for an off-axis spectral index measurement experiment are included.

\end{abstract}


\section{Introduction} \label{sec:intro}

The MeerKAT telescope is a radio interferometer containing 64 antennas separated by up to 8 km and is situated in the Northern Cape Province of South Africa \citep[see, e.g.,][]{Mauch_2020}. Holographic measurements of its 13.5 m diameter antennas have been performed in \em Ku\em -band (11.5 -- 12.5 GHz) on satellite beacon targets as early as 2015. While excellent for collimation checks and dish surface accuracy measurements, such targets are not suitable for measuring full-polarization beam shapes. Not only do satellite targets provide beacons at limited frequencies, these signals are highly polarized which renders the measurement equation \citep{Smirnov_2011} not invertible, and in turn leaves derived full-polarization patterns ill-defined. The high channelization required to observe a narrow band beacon signal could also be wasteful. Furthermore, since there exists only 3 \em Ku\em -band receivers and it is labour intensive to reinstall the units onto subsequent antennas, it is impractical to monitor and comprehensively characterize the 64 antenna array in \em Ku\em -band.

Holography was also performed in \em L\em -band (856 -- 1711 MHz) around the same time to measure beam shapes for qualification purposes. The measurement techniques have been refined during the roll-out, and since 2019 the beam shapes for all antennas have been monitored routinely under different environmental conditions for historical record. Diffraction effects make accurate dish surface and collimation measurements more challenging in \em L\em - than \em Ku\em -band, but the full-polarization beams can be measured rapidly up to a few sidelobes at all frequencies using unpolarized celestial targets. Similar measurements are performed in \em UHF\em - and \em S\em -bands, but these fall outside the scope of this paper.

Of immediate interest to the community is the nature of the MeerKAT primary beams in \em L\em -band, which is its most popular observing band. The telescope was designed with the goal that tolerances be tight enough so that it would not be necessary to model any antenna differently from the array average. Individualizing the beams would have a computational cost implication which is intentionally avoided. Equivalently, a cheaper mechanical structure could have been built if it was not for this design principle. 

Unfortunately design stage fundamental mode electromagnetic simulations did not predict frequency dependent pointing behavior that severely impacts primary beams in the upper half of the band. A recent study \citep{deVilliers_2021} uncovered that higher order waveguide modes are activated by the orthomode transducer (OMT) within the operating band, and are responsible for the effects observed.

This paper communicates measured properties of the primary beams in figures for reference, and quantifies changes in the beam shapes that may occur. As a guide to telescope users, power levels are quoted of effects that would need to be taken into account in order to achieve the required level of accuracy.

\section{Definitions and measurement technicalities}

Only limited descriptions are provided here of the holographic measurement techniques used since this topic is adequately discussed in the literature, e.g. \cite{Perley_2021}, \cite{Perley_2016}, \cite{Popping_2008}, \cite{Harp_2011}, \cite{Iheanetu_2019} and \cite{Asad_2019}. Some details that are peculiar to this work and relevant to telescope science users, are discussed below.

\subsection{Scan pattern}
\begin{figure}[t]
\centering
\includegraphics[width=\linewidth,trim=0.5cm 0.2cm 1.6cm 1.25cm, clip]{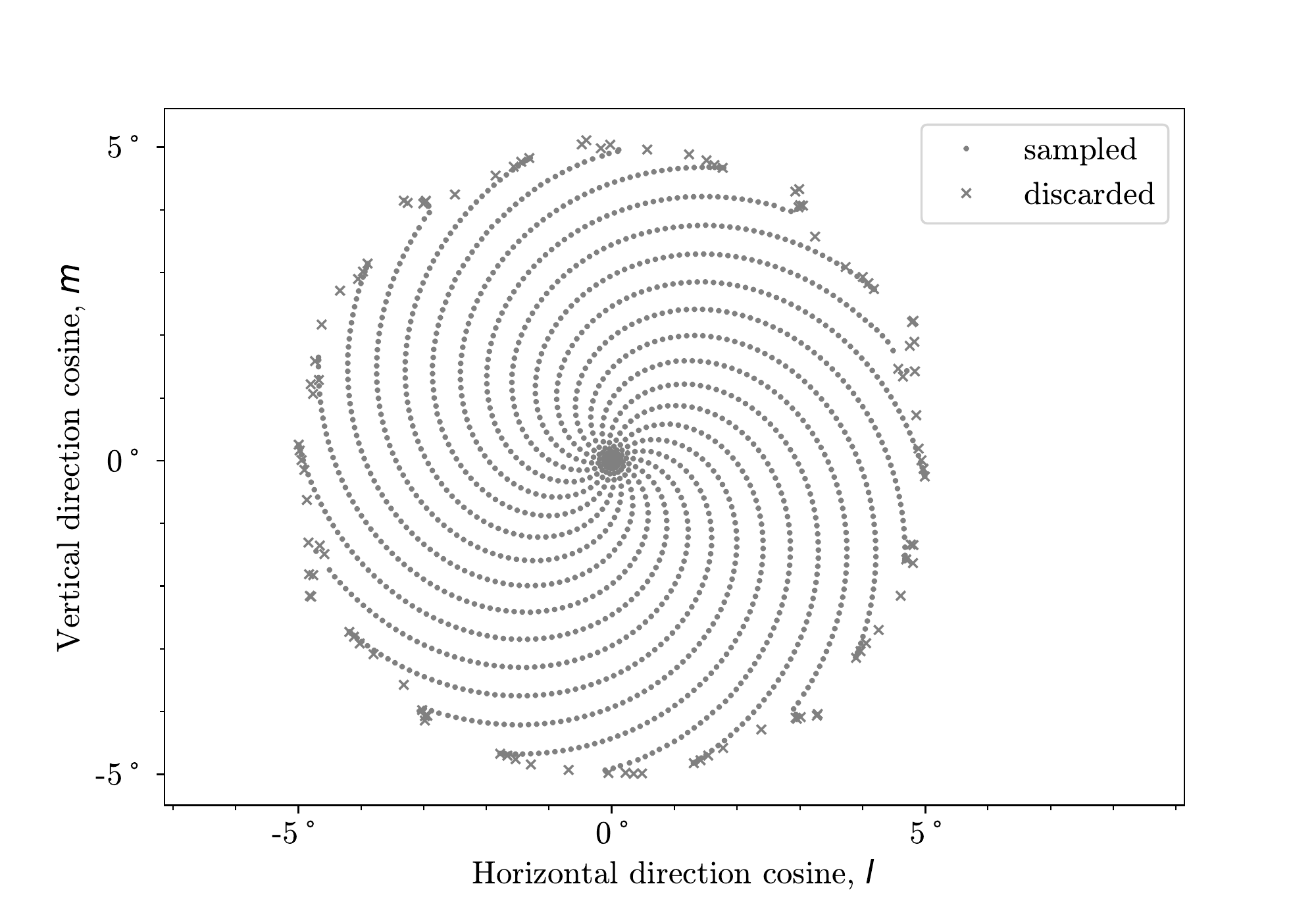}
\caption{A typical spiral scan pattern observed in \em L\em -band showing sampled and discarded slew points.
\label{fig:scanpattern}}
\end{figure}

In the simplest case of a geostationary target, a beam pattern could be sampled by scanning in azimuth for a number of different elevations. However, observation time is squandered on slewing motions because the on-target direction needs to be re-observed regularly for calibration purposes. Even worse, for an observation strategy where multiple rapid pointings are made in a grid, additional settling time allowances are needed at each pointing before capturing usable data. Overheads become costlier when larger angular distances are travelled at low frequencies, for relatively small diameter antennas, or at high elevations. It is more efficient to scan in a spiral pattern where much less data is discarded. A practical spiral pattern as in Figure \ref{fig:scanpattern} scraps about 8 times less data than a raster pattern (with same scan speed, extent and on-target regularity), while achieving scan spacings that are twice as dense. 

A spiral scan pattern is preferred because it reduces the required observation time, has a circular rather than rectangular footprint and offers variable control over the degree of sampling uniformity across the beam. It is useful for wide-band beam measurements to sample the central part of the beam more densely to accommodate the narrowing of the beam width at higher frequencies. With an extreme twist factor of zero, a radial scan pattern would oversample the center and undersample the outer regions of the beam.

Since MeerKAT antennas can deform under environmental conditions it is useful to minimize the observation time in order to distinguish effects especially when scanning celestial targets that move across the sky. The benefit of doing many shorter observations instead of fewer longer duration ones (that could individually attain higher signal to noise ratio) goes beyond isolating environmental effects on the beam shape: radio frequency interference can almost be eliminated given enough time and orientation variability, allowing excellent beam shape recovery over the full band. 

Unlike most observation scripts, the holography script has a short minimum observation time, and has no onerous restriction on minimum number of antennas, which allows it to be scheduled easily and regularly as a filler at any time. Targets are by default chosen automatically to maximize diversity in elevation amongst observations, and to avoid the Sun, when applicable.

Individual observations cycles are 10$^\circ$ in extent, 30 minutes in duration and captured at 1 second dump intervals, with 1024 channels across the band. However, at elevations greater than $60^\circ$ the cycle duration is increased due to further distances that need to be travelled by the azimuth drive. When necessary, the array is unwrapped at the start of holography observations to ensure that each antenna has the same relative range of motion in azimuth. The ordering of the scan pattern arms are chosen to reduce the elevation extent to less than what is possible for geostationary targets.

Usually half the antennas in the array are scanned while the remainder tracks the target for reference. These groups are alternated in subsequent cycles, however, antennas under investigation can be selected to be scanned consecutively.

\subsection{Direction cosine feed plane coordinates}
In order to command the azimuth elevation driven antennas to follow a desired scanning pattern when sampling the beam, feed plane coordinates must be converted to requested azimuth elevation coordinates.

The forward equations for mapping a scanning antenna's measured azimuth and elevation coordinates ($\phi_\mathrm{scan}$,$\theta_\mathrm{scan}$) to the beam plane, are
\begin{equation}
l = \cos(\theta_\mathrm{target}) \sin(\phi_\mathrm{target}-\phi_\mathrm{scan})
\end{equation}
\begin{equation}
\begin{split}
m = -\cos(\theta_\mathrm{target})\sin(\theta_\mathrm{scan})&\cos(\phi_\mathrm{target}-\phi_\mathrm{scan})\\
+ \sin(\theta_\mathrm{target})&\cos(\theta_\mathrm{scan})
\end{split}
\end{equation}
\begin{equation}
n=\sqrt{1-l^2-m^2}
\end{equation}
where $l$, $m$ and $n$ are direction cosines of the scanning antenna's feed orientation with respect to the target at ($\phi_\mathrm{target}$,$\theta_\mathrm{target}$). These equations differ from the standard orthographic projection in that the role of reference coordinate is exchanged with that of the object coordinate. This in turn results in disparate inverse equations,
\begin{equation}
 \phi_\mathrm{scan}=\phi_\mathrm{target}-\sin^{-1}\left(l/\cos(\theta_\mathrm{target})\right)
\end{equation}
\begin{equation}
 \theta_\mathrm{scan}=\sin^{-1}\frac{n\sin(\theta_\mathrm{target})-m\sqrt{\cos^2(\theta_\mathrm{target})-l^2}}{1-l^2}
\end{equation}
which are used to determine the requested coordinates in an observation. Note that at high elevations, scanning must extend further in the azimuthal direction to capture the same beam extent in feed plane coordinates.

\subsection{Feed phase center offset}
A geometric phase difference transpires in the correlation product between two antennas when the orientation of one antenna is rotated relative to that of the other, around a point that is not the antenna's feed phase center.  It is necessary to correct holography visibility data between scanning--tracking baselines for the effect of this phase center offset relative to the antenna mount phase center. The phase adjustment
\begin{equation}
\Delta \Phi_\mathrm{0}={\frac{2\pi}{ \lambda} (\Delta x_\mathrm{0} l+\Delta y_\mathrm{0} m+\Delta z_\mathrm{0} n) }
\end{equation}
is added to the measured visibility data phase where the offset is $\Delta x_\mathrm{0}=0$ sideways, $\Delta y_\mathrm{0}=-1550$ mm downwards, and $\Delta z_\mathrm{0}=-2476.2$ mm backwards, according to design drawings. The adjustment has the effect of centering the aperture around the origin.

\subsection{Errorbeam power metric}

Throughout this paper the errorbeam power metric is used to compare corresponding beam shapes quantitatively against each other. This metric is chosen over the more familiar root mean square (RMS) error because MeerKAT per-antenna primary beams are specified to have an errorbeam power of less than 4\% relative to the array average over the 900 -- 1670 MHz range. The use of this metric makes it easier to see how various effects measure up against this qualification limit.

When comparing beams, the RMS error metric is sensitive to the area size of the beam being evaluated because larger errors tend to exist where the beam has more power. If evaluation is dynamically restricted to the half power region of the beam, then the RMS error estimates an average error over this region while errorbeam quotes the worst case. Because the beam shape errors vary relatively slowly over the half power region, these metrics are about equally informative.

While other definitions exist, in this paper errorbeam power is a function of frequency, is distinguished amongst polarization types $p$, and is expressed in units of percentage power. Errorbeam power, $P_\mathrm{EB}$, is defined as follows:

\begin{equation}
P_{\mathrm{EB},p}(f)=100 \max_{l,m} \left|{|E_p(f,l,m)|^2-|E^\#_p(f,l,m)|^2}\right|
\label{eqn:errorbeam}
\end{equation}
for $|E_I(l,m)|^2>0.5$ only, where $E$ is a complex voltage beam response per feed and polarization type that is being compared to a reference $E^\#$ of the same kind, each of which is normalized to unity maximum of the corresponding copolarization beam. Normally $p$ refers to the horizontal (H) or vertical (V) feed response to a horizontal or vertical polarized wave front, $f$ is frequency, and $l$ and $m$ are direction cosine coordinates of the beam relative to the frequency independent pointing direction of the antenna. For Stokes IQUV errorbeams ($P_{\mathrm{EB},I}$,$P_{\mathrm{EB},Q}$,$P_{\mathrm{EB},U}$,$P_{\mathrm{EB},V}$) the following definitions apply assuming an unpolarized source:
\begin{eqnarray}
|E_{\mathrm{I}}|^2&\triangleq&\frac{1}{2}\left(|E_{\mathrm{HH}}|^2 +|E_{\mathrm{HV}}|^2 +|E_{\mathrm{VH}}|^2 +|E_{\mathrm{VV}}|^2\right)\\
|E_{\mathrm{Q}}|^2&\triangleq&\frac{1}{2}\left(|E_{\mathrm{HH}}|^2 +|E_{\mathrm{HV}}|^2 -|E_{\mathrm{VH}}|^2 -|E_{\mathrm{VV}}|^2\right)\\
|E_{\mathrm{U}}|^2&\triangleq&\Re\left(E_{\mathrm{HH}}\overline{E_{\mathrm{VH}}} +E_{\mathrm{HV}}\overline{E_{\mathrm{VV}}}\right)\\
|E_{\mathrm{V}}|^2&\triangleq&\Im\left(E_{\mathrm{HH}}\overline{E_{\mathrm{VH}}} +E_{\mathrm{HV}}\overline{E_{\mathrm{VV}}}\right)
\label{eqn:stokeserrorbeam}
\end{eqnarray}

It is important to emphasize that some studies remove any pointing errors before drawing comparisons amongst beams when evaluating an errorbeam metric. Great consideration should be given to the appropriateness of whether the pointing error should be included or not when evaluating the errorbeam metric for any given use case because, as seen in Section \ref{section:pointingerror_errorbeam}, this has a dominating impact on the results. 

\subsection{Cosine taper beam model}
\label{section:cosine_taper_beam_model}
The cosine-tapered field pattern, derived in Section 3.2.5 of \cite{Condon_2016}, shares a remarkable resemblance to the MeerKAT beam shape, particularly at 1500 MHz and in the horizontal direction only, moreso than other simple mathematical formulations, as pointed out by \cite{Mauch_2020}. It can be shown that a perfectly cosine-tapered aperture transforms to a voltage beam pattern of the form
\begin{equation}
 E_\mathrm{CT}(r/r_0)=\frac{\cos(\pi r)}{1-4 r^2}
\end{equation}
where $r_0=1.1889647809329453$ ensures that the half power beam width (HPBW) occurs predictably at $E_\mathrm{CT}(0.5)=\sqrt{0.5}$. The publicly available \texttt{katbeam}\footnote{\href{https://github.com/ska-sa/katbeam}{https://github.com/ska-sa/katbeam}} module constructs simplified MeerKAT beam patterns using this function in conjunction with offset ($l_0,m_0$) and beam width ($l_\mathrm{HPBW},m_\mathrm{HPBW}$) measurements as follows:

\begin{equation}
 r=\sqrt{\left(\frac{l-l_0}{l_\mathrm{HPBW}}\right)^2+\left(\frac{m-m_0}{m_\mathrm{HPBW}}\right)^2} 
\end{equation}

In this simple module the Stokes I beam is composed from the copolarization models only and the cross-polarization beams are approximated to zero.

\subsection{Array average concepts}

The array average beam shape may draw great interest, however it is often poorly communicated by telescope users which kind their reduction software require. It may be best to define some options explicitly.

\subsubsection{Half array average}
The beam patterns of only about half of the array is scanned during a typical individual holography observation, because the other half is used for reference. A half array average would differ systematically by some margin from that of the subsequent observation cycle where the alternate set of antennas are scanned. Environmental factors such as changed gravitational loading may further render alternate observation cycles incomparable at some level of accuracy.

\subsubsection{Composed array average}
A full array average can be composed using results from more than one comparable holography observation cycle, by first removing any pointing errors prior to averaging applicable beams together. Technically an array average could be tailored for a particular science observation from only the antennas actually used, at a given elevation and ambient temperature. Most commonly an average is composed from an equal weighting per antenna over the full array at an elevation of 60$^\circ$ and ambient temperature of 15$^\circ$C. Such an average may resemble a high quality idealized result that eliminates antenna specific assembly imperfections, and is referred to as the \em array average \em throughout this paper when not explicitly qualified differently.

\subsubsection{Widened beam array average}
Image plane primary beam corrections are performed after the image has been generated. The applicable primary beam pattern that should be used to normalize the flux across the image would be an average effect of the particular antennas used to produce the image, where the beams are offset, rotated and smeared by the corresponding parallactic angle rotation over the sky. Furthermore, for continuum imaging the applied beam would be an average over the band. The antenna pointing errors, which for MeerKAT are substantial (see Section \ref{section:pointingerror_errorbeam}), should be included in such a computation because it has the effect of broadening the beam wider than that of individual antennas, and also offsets the array average beam. The errorbeam deteriorates more than expected because the pointing error of the array average beam is also large.

Depending on the application, it may be preferable to widen a composed array average beam by convolving it with a pointing uncertainty distribution. Although the factor by which the beam widens is small relative to the beam width in \em L\em -band, it could impact the accuracy of sensitive spectral index calculations. 

\begin{deluxetable*}{ccccccccc}[!t]
\tablenum{1}
\tablecaption{Errorbeam power dependence on measurement SNR for individual holography measurements\label{tab:snr_errorbeam}}
\tablewidth{0pt}
\tablehead{
\colhead{SNR} & \colhead{Target} & \colhead{Flux density} & \multicolumn2c{Stokes I}& \multicolumn2c{Horizontal copolarization} & \multicolumn2c{Vertical copolarization}\\
\colhead{[Ratio]} &\colhead{[Name]} &\colhead{[Jy$_{1.4 \, \mathrm{GHz}}$]} & \multicolumn2c{Errorbeam power [\%]} & \multicolumn2c{Errorbeam power [\%]} & \multicolumn2c{Errorbeam power [\%]} \\
\colhead{} & \colhead{} & \colhead{} & \colhead{median} & \colhead{16-84$^\mathrm{th}$ percentile} & \colhead{median} & \colhead{16-84$^\mathrm{th}$ percentile} & \colhead{median} & \colhead{16-84$^\mathrm{th}$ percentile}
}
\startdata
30  & J1924-2914 & 12 & 0.55 & 0.31-1.63  & 0.62 & 0.35-1.74 & 0.61 & 0.34-1.63  \\
60  & PKS 1934-63 & 15 & 0.29 & 0.17-0.55  & 0.34 & 0.20-0.63 & 0.33 & 0.20-0.60  \\
100 & 3C 273 & 36 & 0.17 & 0.11-0.29  & 0.19 & 0.12-0.33 & 0.20 & 0.12-0.35  \\
\enddata
\end{deluxetable*}
\section{Results}

This section presents measured properties of the MeerKAT primary beams in \em L\em -band and quantifies its variability. As a summary preceding the conclusion, a more technical perspective on the power levels at which common effects start playing a role is also provided. Some larger figures and tables referenced in this section are appended at the end of the article to improve the flow of reading. Datasets and selected data products related to this work will be made available electronically\footnote{\href{https://doi.org/10.48479/s9nh-3s43}{doi: 10.48479/s9nh-3s43}}, and primary beam model updates will be distributed through the public facing \texttt{katbeam} module.

\begin{figure}[b!]
\includegraphics[width=\linewidth,trim=0.5cm 0.2cm 1.6cm 1.25cm, clip]{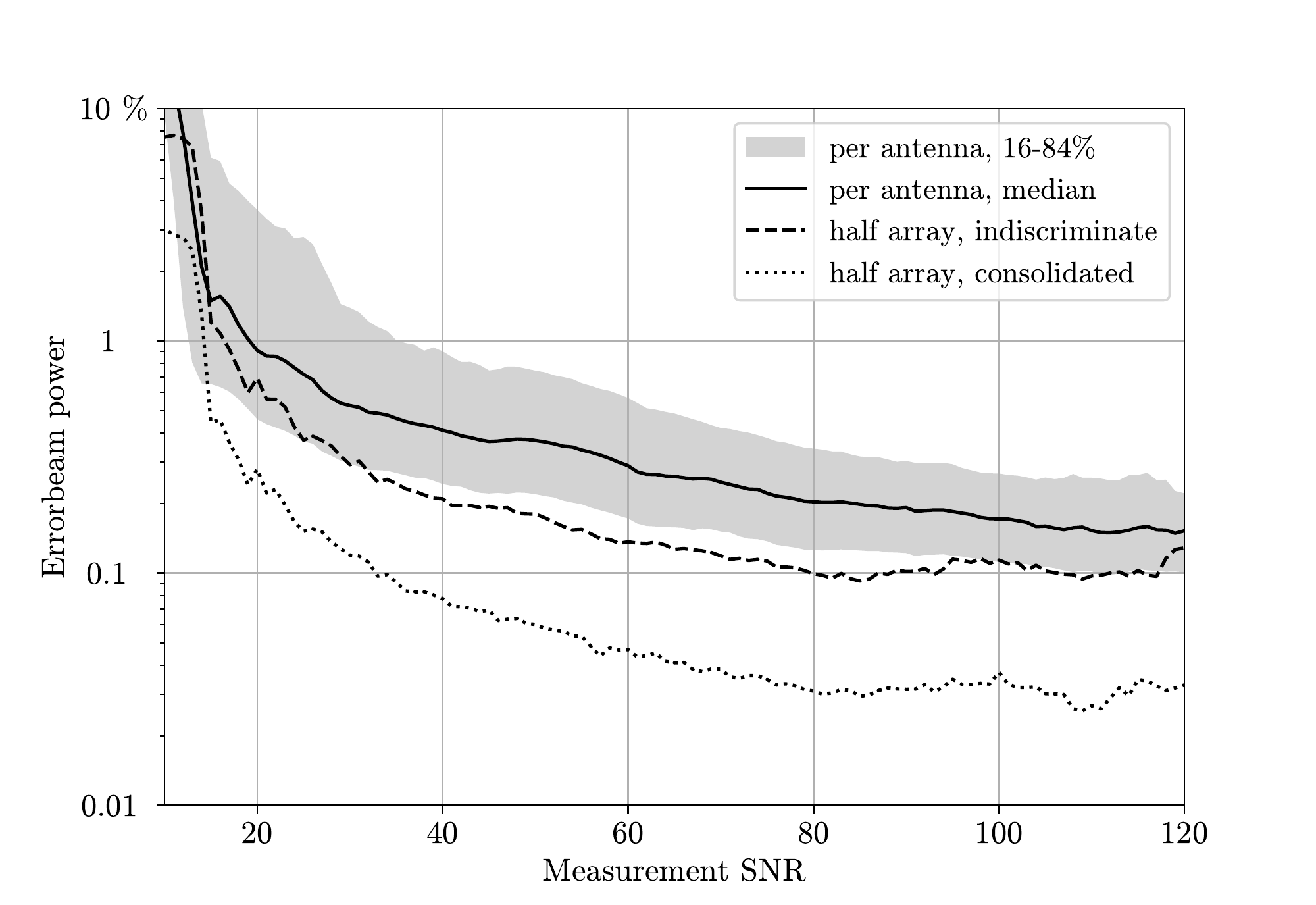}
\caption{Estimated errors in the measured beam shape for individual holography observation cycles are expressed in terms of errorbeam power as a function of the measurement signal to noise ratio.
\label{fig:snr_errorbeam}}
\end{figure}

\subsection{Measurement accuracy}
\label{section:measurement_accuracy}

The accuracy of beam shape measurements depends on the signal to noise ratio of the visibility data, which directly relates to the flux of the target observed. In addition, the scan pattern used, observation duration, missing data, the presence of other targets in sidelobes and radio frequency interference (RFI) also affects the quality of the beam measurement. These circumstantial factors contribute to non-zero Fourier transform components that lies outside the footprint of the aperture. Relying on Parseval's Theorem, the effective signal to noise ratio of a measurement, representative of its overall quality, is estimated from these components in the aperture plane.

Figure \ref{fig:snr_errorbeam} illustrates how errorbeam power relates to the quality of the measurement, per observation. The figure is produced by quoting half the errorbeam power between two successive and comparable beam measurements, indexed by the average aperture plane based SNR estimate for that measurement pair. All measurement pairs are comparable in the sense that beams of the same antenna, with the same receiver installed, is being compared and the observations differ by less than 3$^\circ$ in elevation, and less than 5$^\circ$C in ambient temperature, within a variable yet relatively short period of time. Each channel in the 900 -- 1670 MHz range is evaluated separately, where those affected by RFI, for example, would yield a poor SNR. The spread of results is only illustrated for the per-antenna case to avoid clutter.

Since roughly half (typically 25 -- 32 antennas) of the array is used as scanning antennas per observation cycle, a half array instead of a full array average is also computed for each measurement. If these half arrays are compared \em indiscriminately\em, i.e. different groups of antennas exist in each half array, then the resultant errorbeam is worse than if the results are \em consolidated\em, i.e. half array pairs are composed of the same antennas. The improvement in errorbeam resulting from averaging over multiple ($N$) consolidated scanning antennas seems to follow a $\frac{1}{\sqrt{N}}$ rule.

Table \ref{tab:snr_errorbeam} summarizes the errorbeam power that can be expected in measured beam patterns from individual observations. On average a typical holography observation using a target such as PKS 1934-63 yields a per-antenna beam measurement that is accurate to about 0.29\% in power. Furthermore, the table reads that the error could be as much as 0.55\% within the 84$^\mathrm{th}$ percentile of variability amongst antennas and channels. The flux densities of targets are quoted from \cite{Tingay_2003}.

If half of all available observations (selected at random) are used to produce a composite result, and is compared to another independent composite result using the remaining half of observations, then the per-antenna measurement accuracy in errorbeam reduces to 0.05\% on average, and 0.02\% for the consolidated full array average. It may be that factors relating to artefacts hampers convergence beyond this level of accuracy. The fact that antenna patterns may change over time is also ignored here. This finding which represents the estimated accuracy to which the beams are known, is quoted later in Table \ref{tab:snr_errorbeam_effects}.

\subsection{The effect of pointing error on errorbeam}
\label{section:pointingerror_errorbeam}

\begin{figure}[t!]
\centering
\includegraphics[width=\linewidth,trim=0.5cm 0.2cm 1.6cm 1.25cm, clip]{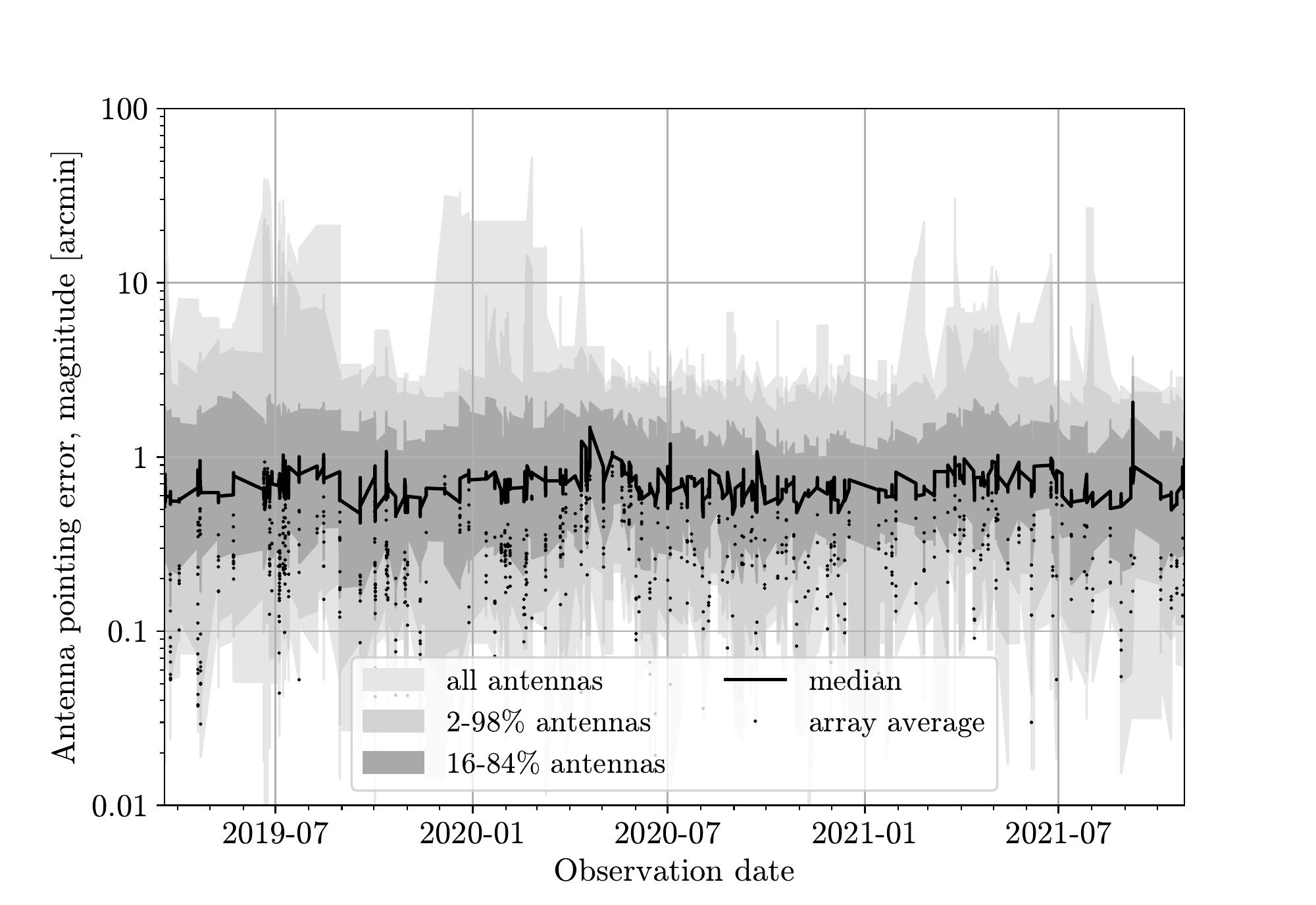}
\caption{An historic record of antenna pointing errors measured by means of holography shows occasional large errors for few antennas. Pointing errors of array average beams are shown using dots. Antenna pointing errors are determined by the weighted fit of phase gradients across the aperture plane. See also Figure \ref{fig:pointing_error} for more details.
\label{fig:antpointing_history}}
\end{figure}

Pointing errors occur when a beam is assumed to be steered precisely towards a certain direction, but in reality it is not. Despite routine pointing calibrations, the MeerKAT telescope operates on a $\sigma=0.64$ arcminute uncertainty in its pointing accuracy. As seen in Figure \ref{fig:antpointing_history}, individual antenna pointing errors are seldom worse than 4 arcminutes, but errors up to a degree have been detected. Reference pointing, where a small update is made based on a nearby target immediately before a target field is observed, is hardly ever used during science observations due to onerous scheduling. While it has shown some improvement in repeatability tests, it is still unclear how well it improves absolute alignment amongst different antennas. System level work on tilt meter sensors is currently being done with the goal to improve pointing accuracy, although simplifying assumptions in system pointing calibration algorithms and outstanding fine-tuning also plays a role. 

While some authors \citep{Cotton_2021} recognize the importance of accurate pointing, it is nevertheless common to assume that the pointing errors are negligible.  Science users have limited means to detect pointing errors from their observations, and if left unaccounted, seemingly small pointing errors could result in intolerable beam shape errors near the half power point of the beam in wide-field imaging projects. Such misalignment may render the need for high accuracy beams useless.

\begin{figure}[t!]
\centering
\includegraphics[width=\linewidth,trim=0cm 0 0cm 1.3cm, clip]{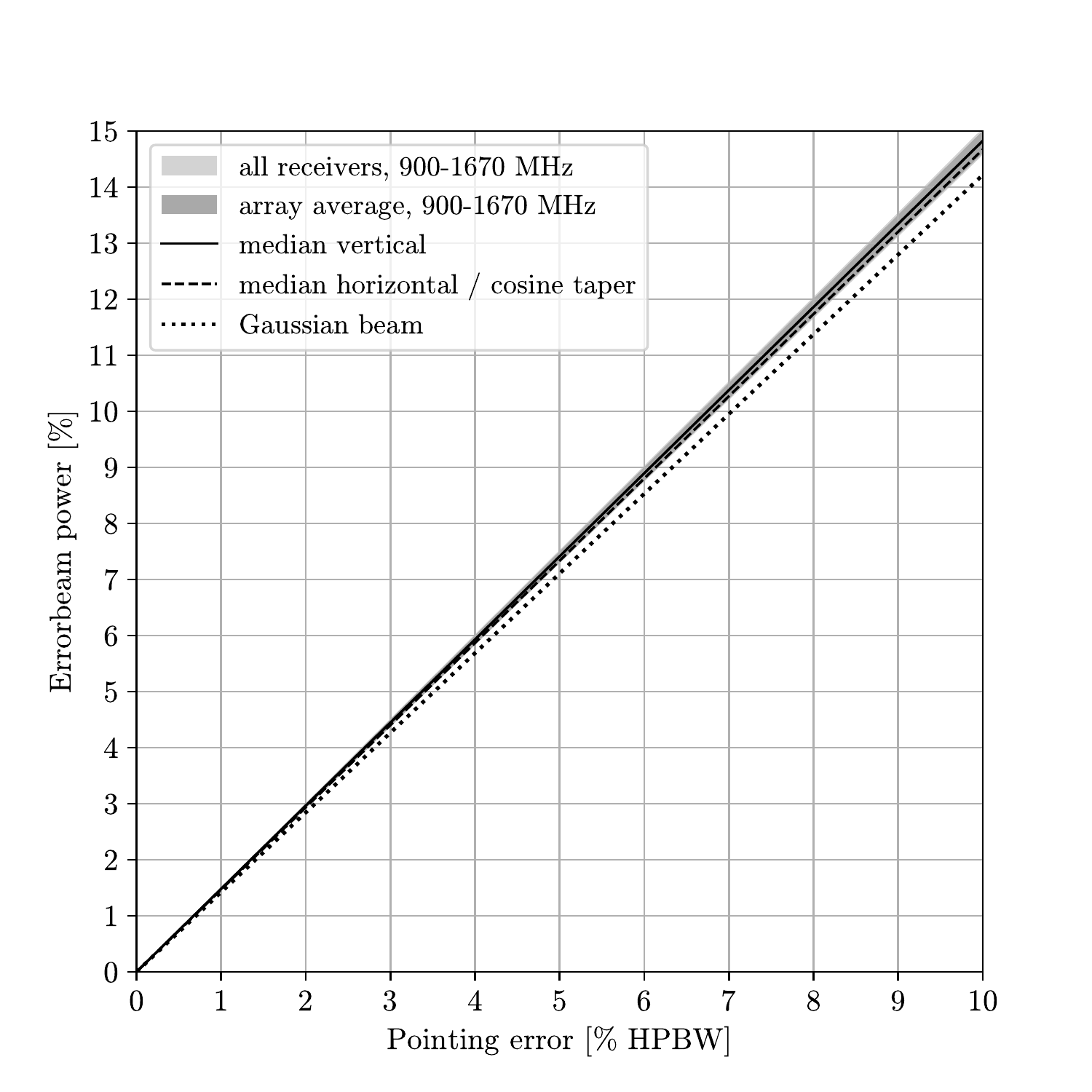}
\caption{The hefty dependence of errorbeam on unaccounted pointing errors cannot be understated as it directly impacts the effectiveness of using high accuracy beams. The peak error in power due to a pointing error occurs in the vicinity of the half power point of the beam.
\label{fig:pointing_errorbeam}}
\end{figure}

The effect of pointing errors on errorbeam can readily be computed using Equation \ref{eqn:errorbeam} for arbitrary beam shapes by offsetting the beam, e.g. $E_p(f,l,m)=E^\#_p(f,l+\Delta l,m)$. If the applied offset is expressed in terms of the half power beam width then the frequency dependence is almost eliminated, and the effect on errorbeam can be graphed compactly as in Figure \ref{fig:pointing_errorbeam}. Barely distinguishable from the performance of a cosine taper beam shape, on average, the Stokes I errorbeam dependence is $1.47\pm 0.02$ times the horizontal pointing error, when expressed as a percentage of the half power beam width. Likewise, the average Stokes I Errorbeam dependence is $1.49\pm 0.02$ times the vertical pointing error percentage. Because the measured beam changes in shape over frequency, and not only in scale, the narrow margin of variation is also illustrated.

Beam shapes with broader peaks relative to the beam width, and hence steeper falloff at the half power point, have a worse errorbeam dependency on pointing error. Results for a Gaussian shaped beam is included for reference. The strong dependency of errorbeam on pointing errors (for any beam shape) plays a dominant role in the usage of high accuracy beam shapes, and is therefore of significant importance. This is shown later in Table \ref{tab:snr_errorbeam_effects}.

\subsection{The half power beam width}

\begin{figure*}[th!]
\centering
\includegraphics[width=\linewidth,trim=2.1cm 0 2.9cm 0, clip]{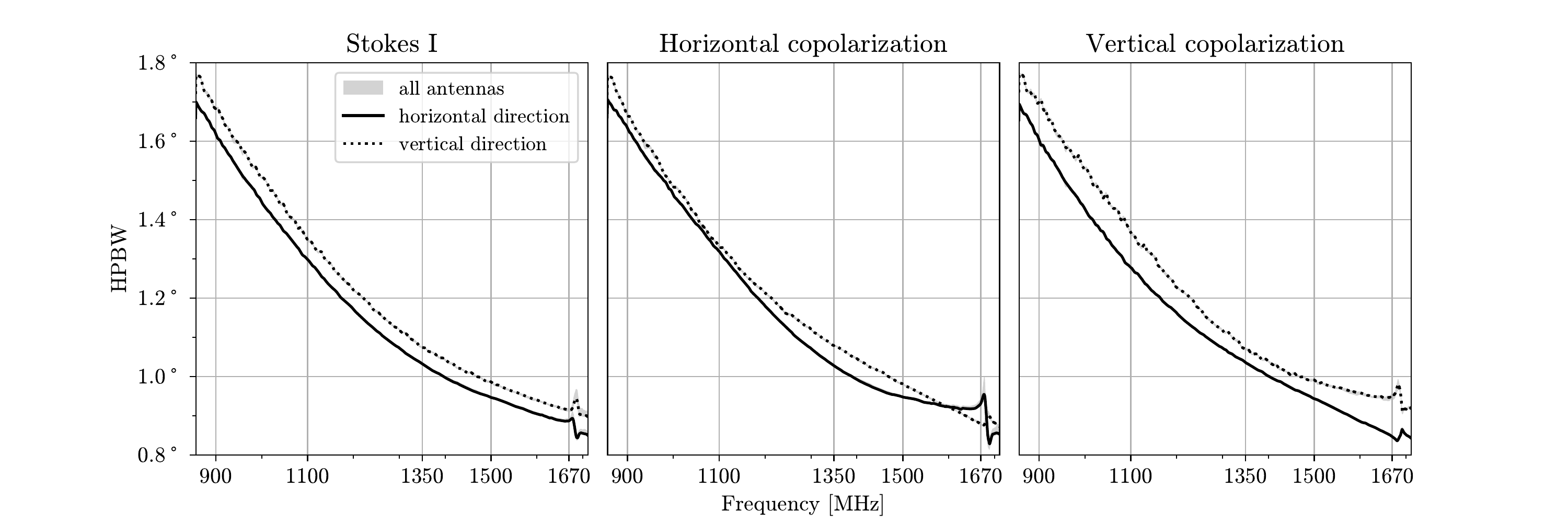}
\caption{The measured half power beam width. \label{fig:hpbw}}
\end{figure*}

Figure \ref{fig:hpbw} shows the measured half power beam width (HPBW) of the copolarization beams for the MeerKAT antennas expressed in degrees as a function of frequency. Results for the Stokes I beam are essentially midway between that of the horizontal and vertical copolarizations. Throughout this paper certain angular distance quantities are conveniently expressed as a percentage of the Stokes I measured HPBW in order to normalize the scale across frequency. There is very little variation amongst different antennas in their half power beam widths: these are within $\sigma=0.1$\% of the HPBW. 

Note that HPBW equals the more familiar full width at half maximum (FWHM) when referring to a power quantity, which is the case here. Reference is also made to half width at half maximum (HWHM = $\frac{1}{2}$ FWHM) in some figures to express radial offset more intuitively.

\subsection{Frequency dependent pointing}

In addition to antenna pointing errors, MeerKAT beams exhibit frequency dependent pointing errors, that also differ per polarization (referred to affectionately as \em squint\em ). This means that if an antenna is optimally lined up to receive maximum power and least cross-polarization impurity from a target at one frequency, for one polarization, it will not be optimally lined up at other frequencies nor for the other polarization. 

These squint profiles as a function of frequency relate to the OMT of the receiver, rather than the antenna onto which the receiver is installed. These squint measurements are repeatable, and are insensitive to collimation effects but deteriorate in the presence of radio frequency interference, degraded calibration due to poor signal to noise ratio (SNR), or data loss effects. Observing the squint is an excellent way to diagnose manufacturing issues internal to the OMT. 

\begin{figure*}[h!]
\centering
\includegraphics[width=\linewidth,trim=2.3cm 1cm 2.9cm 1.5cm, clip]{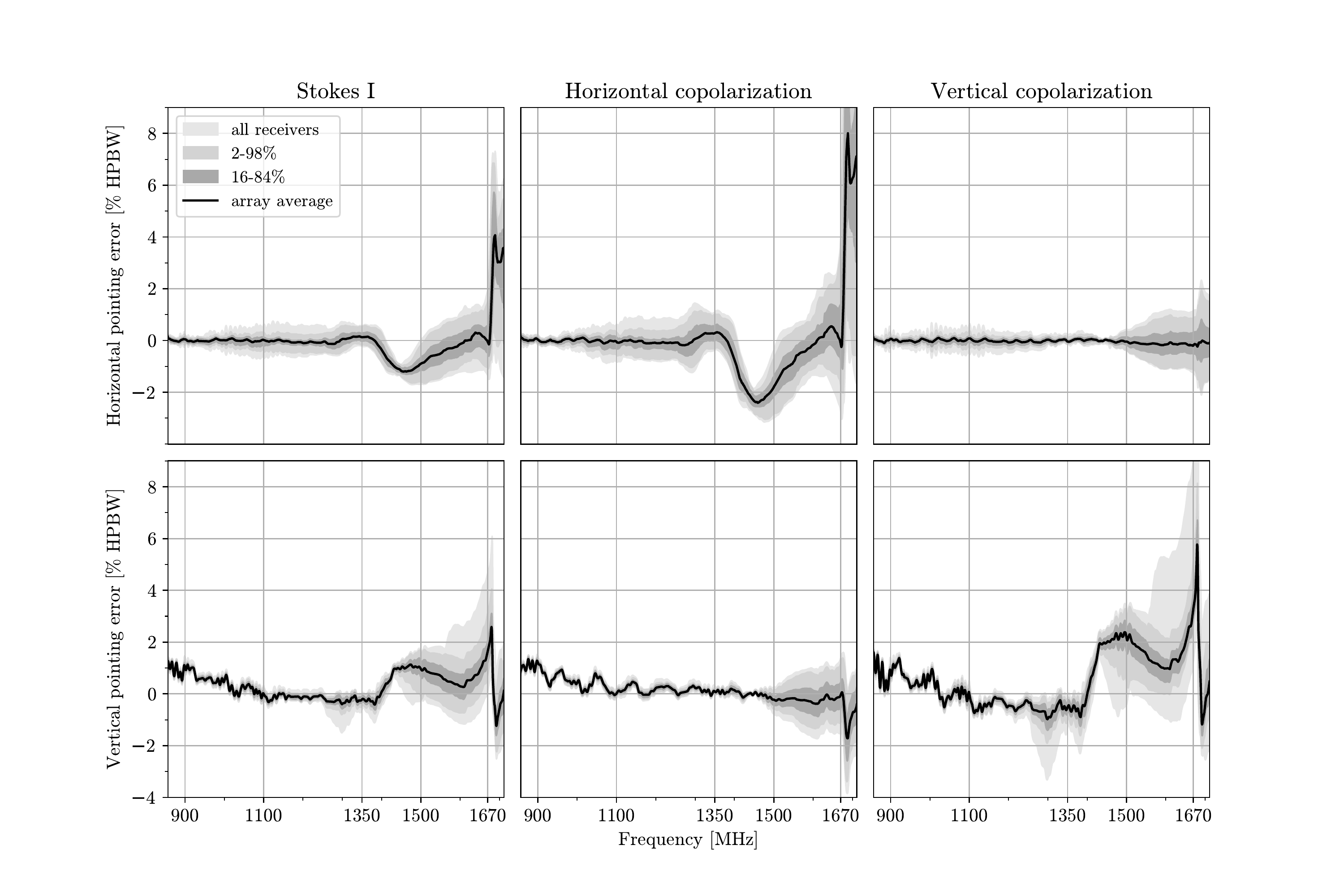}
\caption{Frequency and polarization dependent pointing errors are observed in measured primary beam patterns. The range of variation amongst different receiver units are significant in the upper half of the band. These profiles are independent of the antenna onto which the receiver is installed. Antenna pointing is aligned at 1420 MHz for \em L\em -band, in the horizontal direction using the vertical copolarization beam, and in the vertical direction using the horizontal copolarization beam. \label{fig:pointing_error}}
\end{figure*}

Figure \ref{fig:pointing_error} summarizes measured frequency dependent pointing profiles for all of the 70 possible receivers that might be installed (there are 6 spares). Significant pointing errors exceeding 2\% of the half power beam width at 1500 MHz is noted in the array average for the copolarization beams relative to the nominal pointing direction. These would result in more than 3\% errorbeam power near the half power point if the beam is naively assumed to be centered around the pointing center instead of being offset as a function of frequency as indicated in the figure. The situation is worse at the uppermost end of the band. Although the trend follows the array average, deviations can be substantial for outlier receivers like l.4046 which is responsible for the case that draws attention in the vertical copolarization results.

As for traditional pointing errors, the effect of frequency dependent pointing errors on errorbeam power can be overwhelming. Therefore it is important that any reduction software that relies on the beam shape be mindful of this characteristic, perhaps moreso than the shape itself. In fact, MeerKAT's poor system pointing performance is in part due to the fact that pointing calibration algorithms currently are agnostic to the frequency dependent pointing behavior.

\subsection{The measured copolarization beams}
\label{section:the_measured_copolarization_beams}
Measured MeerKAT beam shapes have been analyzed by fitting elliptical contours through mainlobes at various power levels. Doing so reveals center offsets, semi-major and minor axes as well as orientation angles. Of immediate interest are the results at the half power point, which can be used to normalize a measured beam geometrically into a radial function, approximately.

\begin{figure}[t!]
\includegraphics[width=\linewidth,trim=0.5cm 0.2cm 1.9cm 1.25cm, clip]{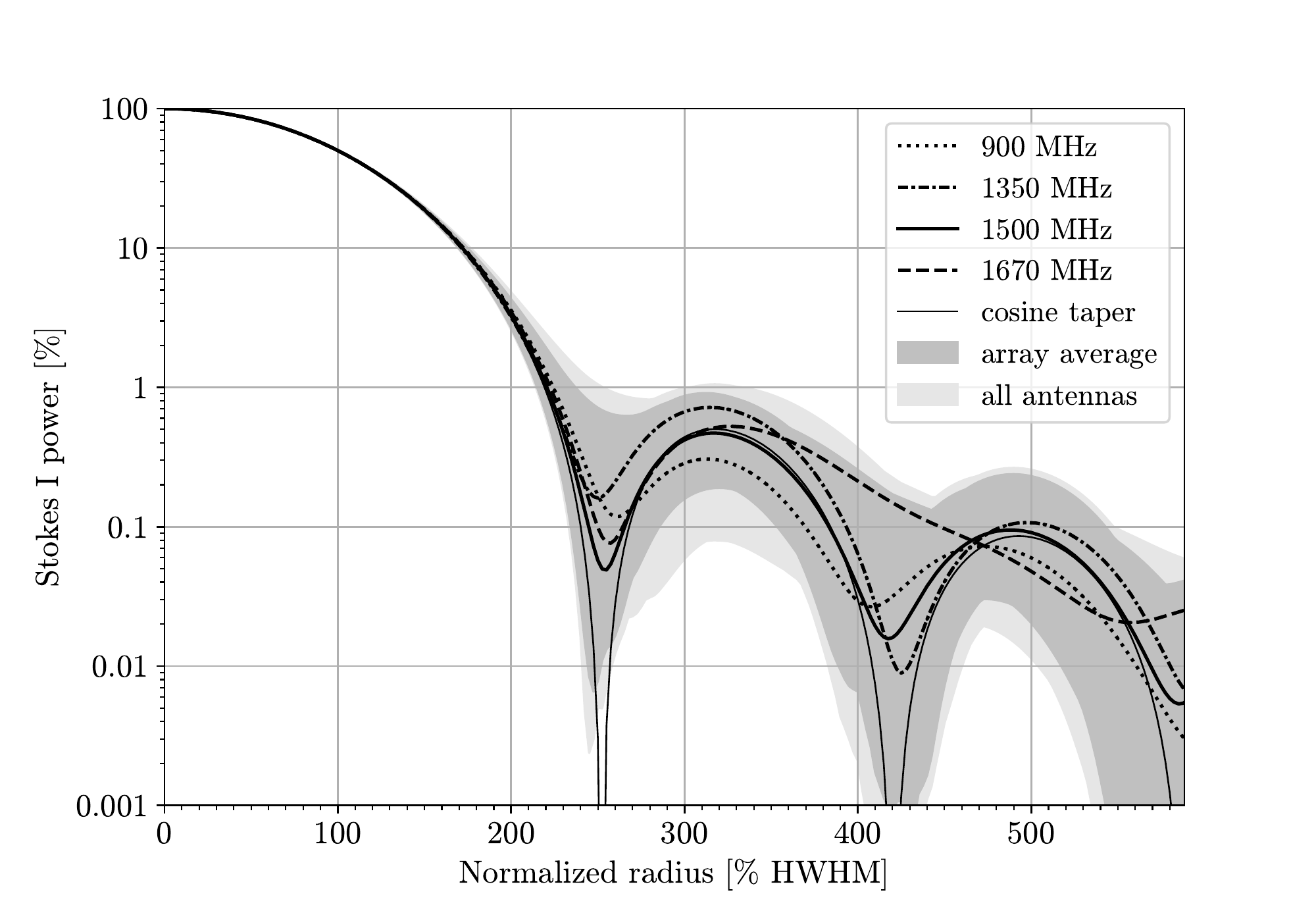}
\caption{This figure shows radial profiles of measured Stokes I primary beams in \em L\em -band that have been geometrically normalized by translation, rotation and radial scaling using the parameters of the best fit ellipse through the half power point of the main lobe. The average value over tangential angle for selected frequencies are indicated using black lines. Despite variations in the sidelobe levels of the profiles at different frequencies, the mainlobe remains remarkably stable after the geometric normalization procedure, and can be well modeled using the cosine taper function. Note that at 1500 MHz, the first sidelobe almost matches the cosine taper function, and at 1670 MHz, the second sidelobe cannot be distinguished from the first sidelobe. The darker grey envelope indicates the extent of variation over the 900--1670 MHz bandwidth and full-polarization angle of the array average Stokes I beam. The lighter grey envelope shows a small widening due to antenna disparities. Results in this figure are for 60$^{\circ}$ elevation and 15$^{\circ}$C ambient temperature.  \label{fig:radial_stokes_I}}
\end{figure}

As illustrated in Figure \ref{fig:radial_stokes_I}, the radial profile of the geometrically normalized measured MeerKAT beam shape at 1500 MHz resembles the cosine taper. Even somewhat beyond the half power point, for all frequencies, this simple mathematical form seems to approximate measurements well at this scale. However, over the range of frequencies in the band and at different tangential locations in the beam, the cosine taper may misrepresent the first sidelobe by about 0.5\% of the on-axis power, which equals the sidelobe power level of this cosine taper itself.

\begin{figure}[t!]
\includegraphics[width=\linewidth,trim=0.5cm 0 1.9cm -0.4cm, clip]{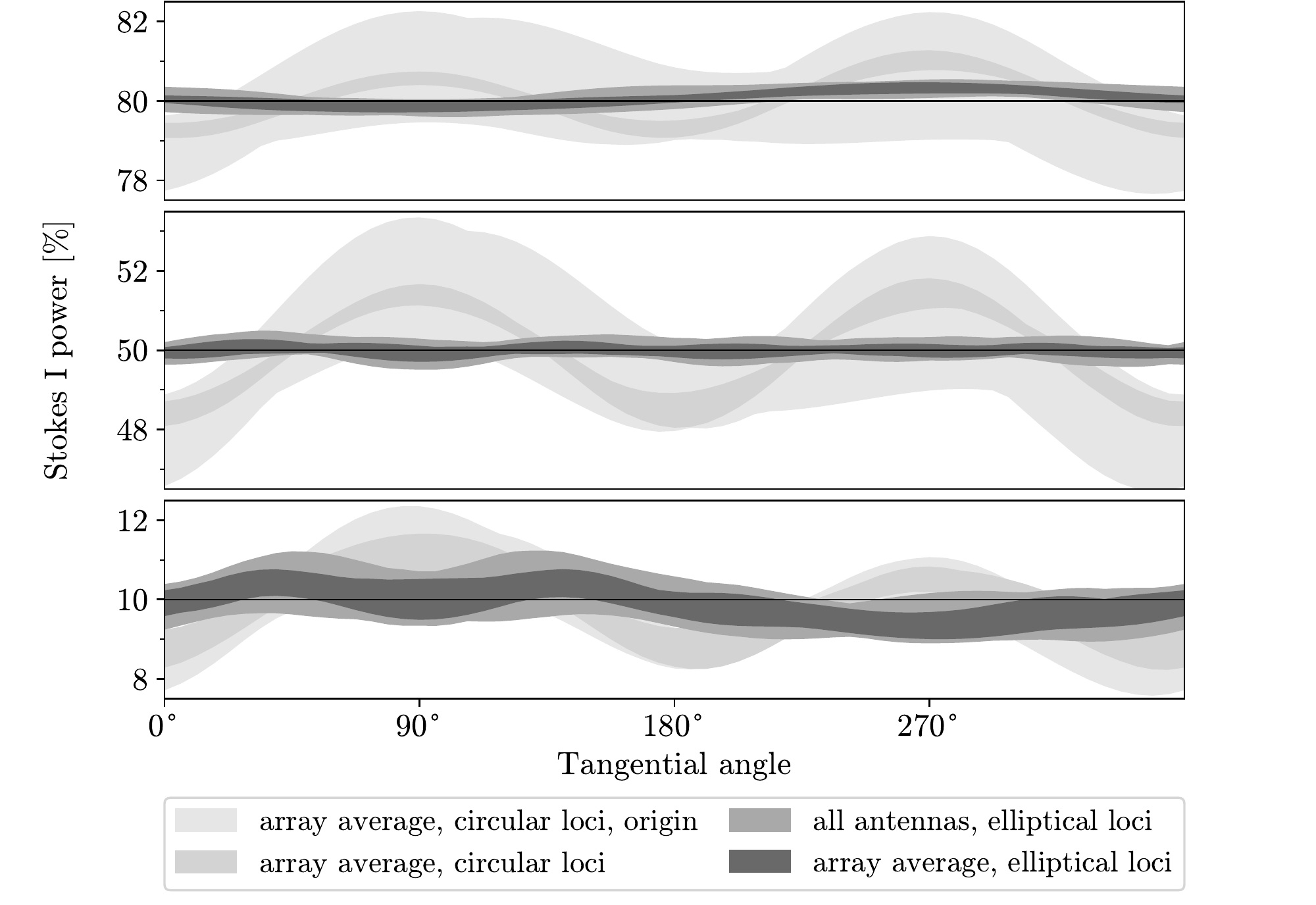}
\caption{This figure shows variations in the power of the Stokes I measured beam around elliptical loci that are best fit to the half power point contour. Three panels are shown corresponding to the 10\%, 50\% and 80\% power levels. The envelope thicknesses are due to variations over the 900--1670 MHz bandwidth, and not due to measurement inaccuracy. A straight flat line at 0.5 cannot be achieved because the half power point beam contours are not perfectly elliptical. A narrow envelope (0.5$\%$ peak to peak) is found for the array average case using an elliptical contour, and only a modest widening occurs (0.9$\%$ peak to peak) when considering individual antenna peculiarities. A significant deterioration (3.7$\%$ peak to peak) is observed if a circular contour is assumed for the Stokes I beam, and an even greater deterioration (6.8$\%$ peak to peak) occurs if frequency dependent centering is not done.
\label{fig:tangential_stokes_I}}
\end{figure}

Tangential cuts through the Stokes I mainlobe are shown in Figure \ref{fig:tangential_stokes_I}, along the loci of best fit ellipse contours to the 10\%, 50\% and 80\% power levels. Cuts along best fit circular loci, including circular loci around the pointing direction (origin) instead of the center offsets individually best fit per frequency channel, are also shown. The significant fluctuations above and below the 50\% power level along the circular loci shows that a naive assumption that the beam is circular is false, and would result in a $\pm 2$\% error in power estimation, which would be increased to $\pm 3$\% if it is also assumed that the beam centering does not change with frequency. Fluctuations along 50\% power elliptical contours are much smaller indicating that the shape is quite elliptical. The 360$^\circ$ tangential period fluctuation along the 80\% power elliptical contours is due to a slight vertical shift of the beam centering, relative to that of the 50\% power elliptical contour. The same phenomenon is seen as fluctuations in the opposite direction, at the 10\% level. Further departures at the 10\% level suggests that the contours appear slightly more triangular.

Although not explicitly depicted in another figure due to its subtle nature, the mainlobe has a skewness in the vertical direction such that the beam centering of the peak power differs by about 2\% of the half power beam width from the centering at the 10\% power level; and by about 0.5\% at the half power level. This is due to the offset Gregorian design of the reflector optics, and is consistent with simulations.

\begin{figure*}[h!]
\includegraphics[width=\linewidth,trim=2.3cm 0 2.9cm 0, clip]{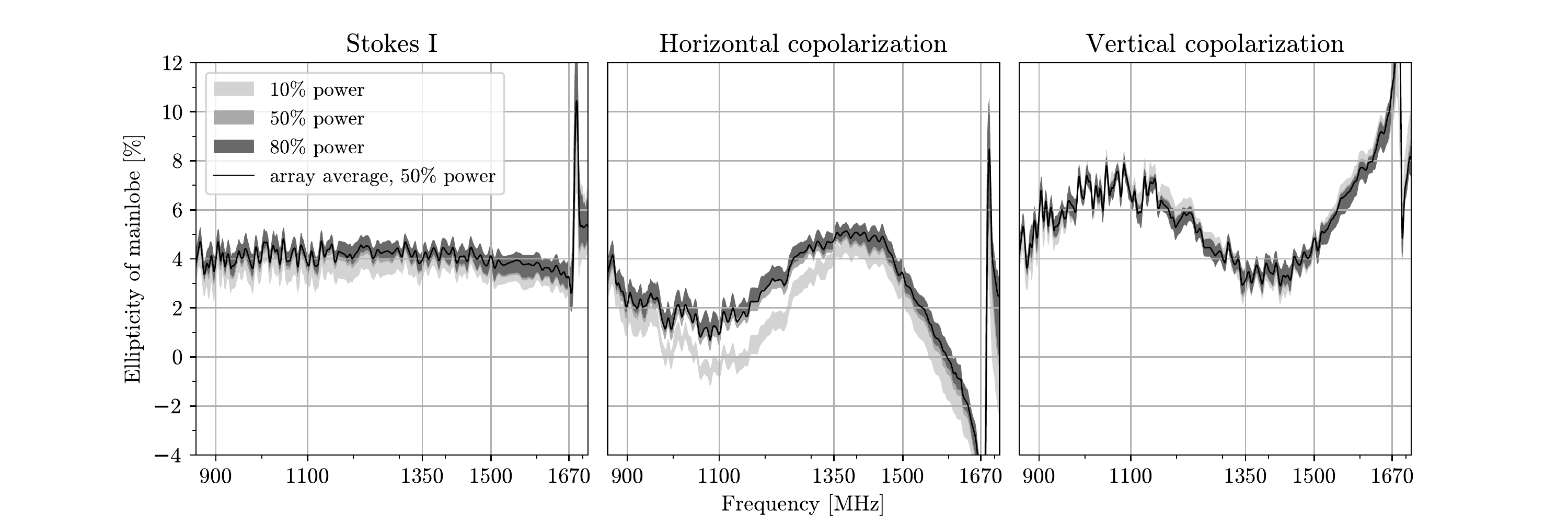}
\caption{The ellipticity of MeerKAT beams at 60$^{\circ}$ elevation, 15$^{\circ}$C are shown at three power contour levels, with envelopes that indicate little variability over all antennas.
\label{fig:ellipticity}}
\end{figure*}

Due to a horizontal broadening of the main reflector design, the Stokes I primary beam is elongated vertically with approximately 4\% ellipticity, as shown in Figure \ref{fig:ellipticity}. At the half power level, the horizontal polarization is elongated horizontally around 1550--1670 MHz only, indicated using negative ellipticity values in this figure. Some further elongation in the horizontal direction occurs at low power levels of the horizontal copolarization mainlobe. Little variation in ellipticity is observed amongst different antennas. 

The rotational orientation of the Stokes I beams for the antennas are within $\sigma=1.8^\circ$ from the axes, except for a few outliers. There is a strong correlation between feed indexer alignment and copolarization power beam orientation, which differs from the actual dipole orientation discussed in the next subsection. The indexer encoder mountings for antennas M023, M024, M033 and M054 have slipped significantly over time. In the worst case, the indexer slippage of 3.2$^\circ$ over $2\frac{1}{4}$ years resulted in a 19$^\circ$ rotation of the Stokes I beam on M054. This rotational effect on the beam tapering also stems from the offset Gregorian reflector optics design.

Telescope users should be aware of a far out sidelobe approximately 72$^\circ$ overhead from the boresight direction at the 0.01\% power level. This sidelobe can pick up interference when it unexpectedly aligns towards a satellite, or the Sun during daytime. At high elevations this lobe is known to pick up terrestrial interference. 

\subsection{The measured cross-polarization beams}

The measured MeerKAT cross-polarization beam patterns agree fairly well with design stage simulations in the lower half of the band only. In the upper half of the band systematic features appear that breaks the symmetry which is presupposed by the reflector optics design.  Even with perfect reflector collimation and despite meticulous symmetry of the feed horn and dipoles within the feed, asymmetry exists in beam patterns above $1284$ MHz. This effect is consequent to the activation of higher order waveguide modes within the OMT that extends behind the dipoles in the receiver unit. The OMT contains small, unavoidable asymmetries that can only be compensated for electromagnetically at limited frequencies, and this causes activation of the TE21 asymmetric mode, increasingly noticeable above its theoretical cutoff frequency of 1387 MHz for the given waveguide diameter. However MeerKAT's wideband specification up to 1670 MHz extends well beyond the cutoff frequency of both the TE21 and TM11 modes resulting in these two modes significantly affecting the beam patterns in half of the operating range of the antennas.

\begin{figure*}[ht!]
\includegraphics[width=\linewidth,trim=0.0cm 0 .0cm 0, clip]{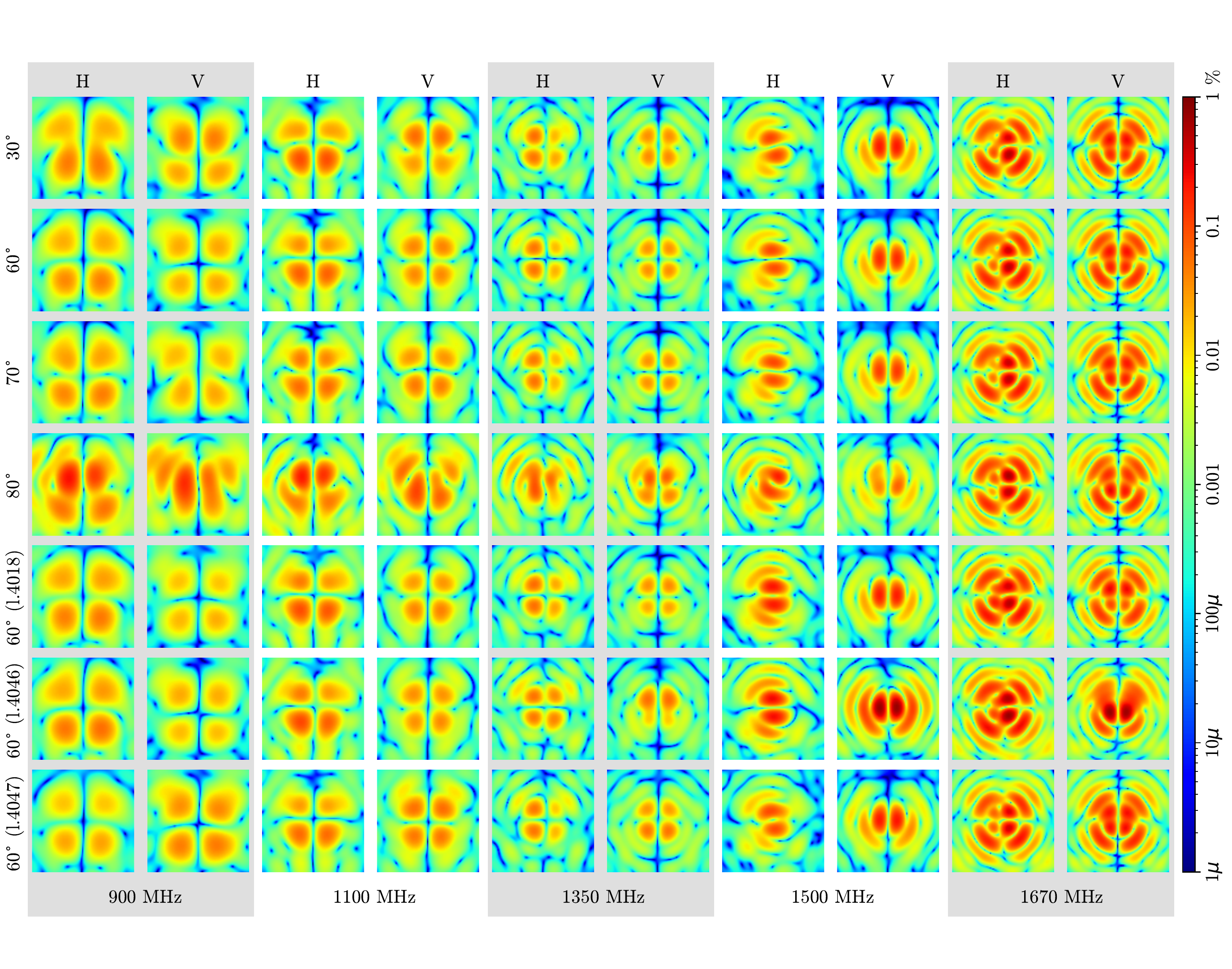}
\caption{Cross-polarization beam shapes are shown for the array average at selected frequencies and elevation angles in the first 4 rows. In the last 3 rows, results for selected receivers at 60$^\circ$ elevation are illustrated. Below 70$^\circ$, gravitational loading affects cross-polarization beam shapes at the 0.05\% power level across the full band, which is at the same or greater level than what receiver to receiver differences impact the lower half of the band (below $1284$ MHz). In the upper half of the band, however, elevation effects are hardly noticeable against the backdrop of more dominant receiver specific effects. Receiver l.4046 fails specifications at some extremes. Each panel shows an 8$^\circ$ extent of the power response of H and V linear feeds due to excitation by an incoming linearly polarized plane wave of the complementary (i.e. V, H respectively) linear polarization. The color scale is relative to the maximum copolarization response.
\label{fig:cross_img_more}}
\end{figure*}

While the antenna to antenna variations in the cross-polarization beam shapes could be explained by collimation effects in the lower half of the band, this is not true in the upper half of the band. Because the OMT mode effects are much greater than the collimation effects at higher frequencies, antenna to antenna variations in these beam patterns are associated with the receiver installed, rather than the antenna onto which it is installed. The higher order waveguide modes are very sensitive to manufacturing accuracy, which means that receivers are identifiable by the cross-polarization beam pattern they produce irrespective of on which antenna they are installed. While both have an impact, the effect of collimation is dwarfed by that of the OMT in the upper half of the band. This can be seen qualitatively in Figure \ref{fig:cross_img_more} and is expressed more quantitatively later in the next subsection. It does not make sense to use per-antenna beam patterns, unless they are dated, because receivers are exchanged from time to time.

Independently of the copolarization power beam rotational orientation which is determined largely by collimation effects and is therefore antenna specific, the cross-polarization pattern is rotated by the dipole orientation and is receiver specific. The dipole assembly orientation error in the OMT can be estimated by fitting a hyperbolic function to the cross-polarization pattern in the vicinity of the zero crossings, over the frequency range that is unaffected by higher order waveguide modes. The angle of the hyperbole asymptotes follows the rotation error. For MeerKAT, the dipoles are on average aligned to the axes within $\sigma=0.7^\circ$, except for receiver l.4002, installed on antenna M006 since August 2019, which is in error by about 3$^\circ$. Receiver specific features follow when a serviced receiver is installed onto another antenna. Both the dipole orientation and the copolarization beam orientation will jointly affect high precision wide-field polarimetry, however the errors are zero mean and small, so will not impact common primary beam correction strategies that apply corrections after the image is formed.

\subsection{Power levels of errorbeam, sidelobes and cross-polarization}
\label{section:power_levels_of_errorbeam}
\begin{figure*}[h!]
\includegraphics[width=\linewidth,trim=2.5cm 1cm 2.9cm 1.5, clip]{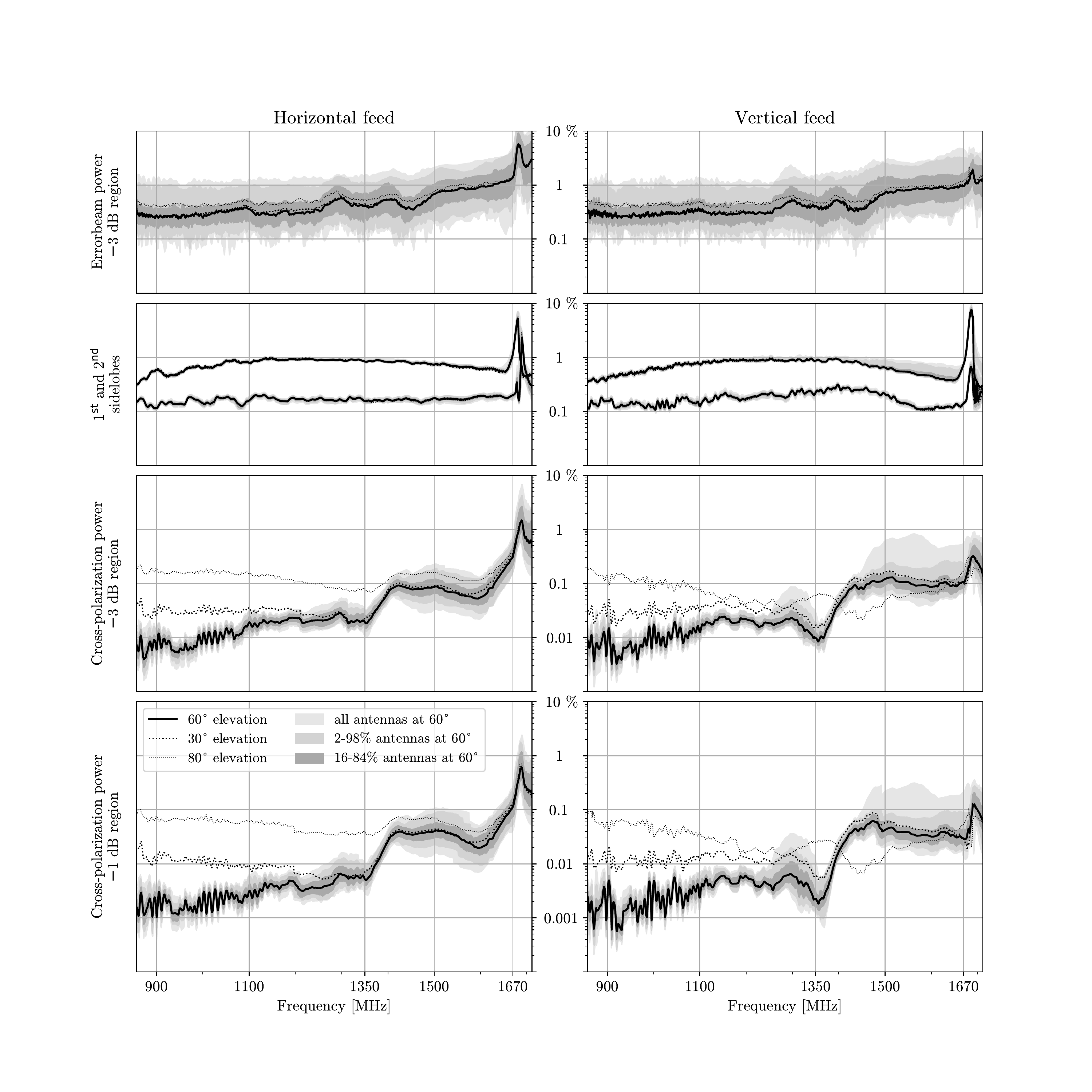}
\caption{The levels of power in antenna to antenna variation in the copolarization beams are compared against sidelobe levels and cross-polarization power, as a function of frequency.
\label{fig:error_freq}}
\end{figure*}

Figure \ref{fig:error_freq} shows the power levels of errorbeam in comparison to sidelobe, and cross-polarization levels, as a function of frequency, for the horizontal and vertical feeds in separate columns. Here the errorbeam is calculated for all antennas, relative to the array average at 60$^\circ$ elevation and 15$^\circ$C.

The horizontal copolarization errorbeam for various antennas differ most commonly from the array average by 0.9\% in power in the 1500--1670 MHz range. Within the 16--84$^\mathrm{th}$ percentile group (equivalent to one standard deviation), antennas differ by up to 1.5\%, and in the worst case up to 5\% in this frequency range. In contrast, at the lower half of the band, antennas differ from the array average most commonly by only 0.3\%, and 0.5\% within the 16--84$^\mathrm{th}$ percentile group, and at worst 1.4\%. The copolarization beam can therefore be approximated 3 times more reliably by the array average in the lower half, than in the upper half of the band. 

Except for the first sidelobe of the vertical copolarization in the upper half of the band, there is little variability amongst different antennas in their peak sidelobe levels, compared to its frequency dependence. The first sidelobe level reaches about 1\% at the middle of the band, and tapers off to about 0.4\% towards both ends of the band. For the horizontal polarization the second sidelobe level is fairly consistently about 0.2\%, whereas for the vertical polarization this ranges between 0.1--0.3\%.

The cross-polarization power levels have a great dependence on frequency as well as elevation, the latter being especially apparent at the lowest end of the band. Per-antenna deviations from the array average is small in the lower half of the band, and much larger in the upper half of the band. The array average makes a better approximation for each beam in the lower half of the band than the upper half. For the lower half of the band it would be more sensible to make provisions for the gravitational loading effect on the cross-polarization pattern, rather than per-antenna differences.

\begin{figure*}[ht!]
\centering
\includegraphics[width=0.78\linewidth,trim=2.5cm 1cm 2.8cm 1.8cm, clip]{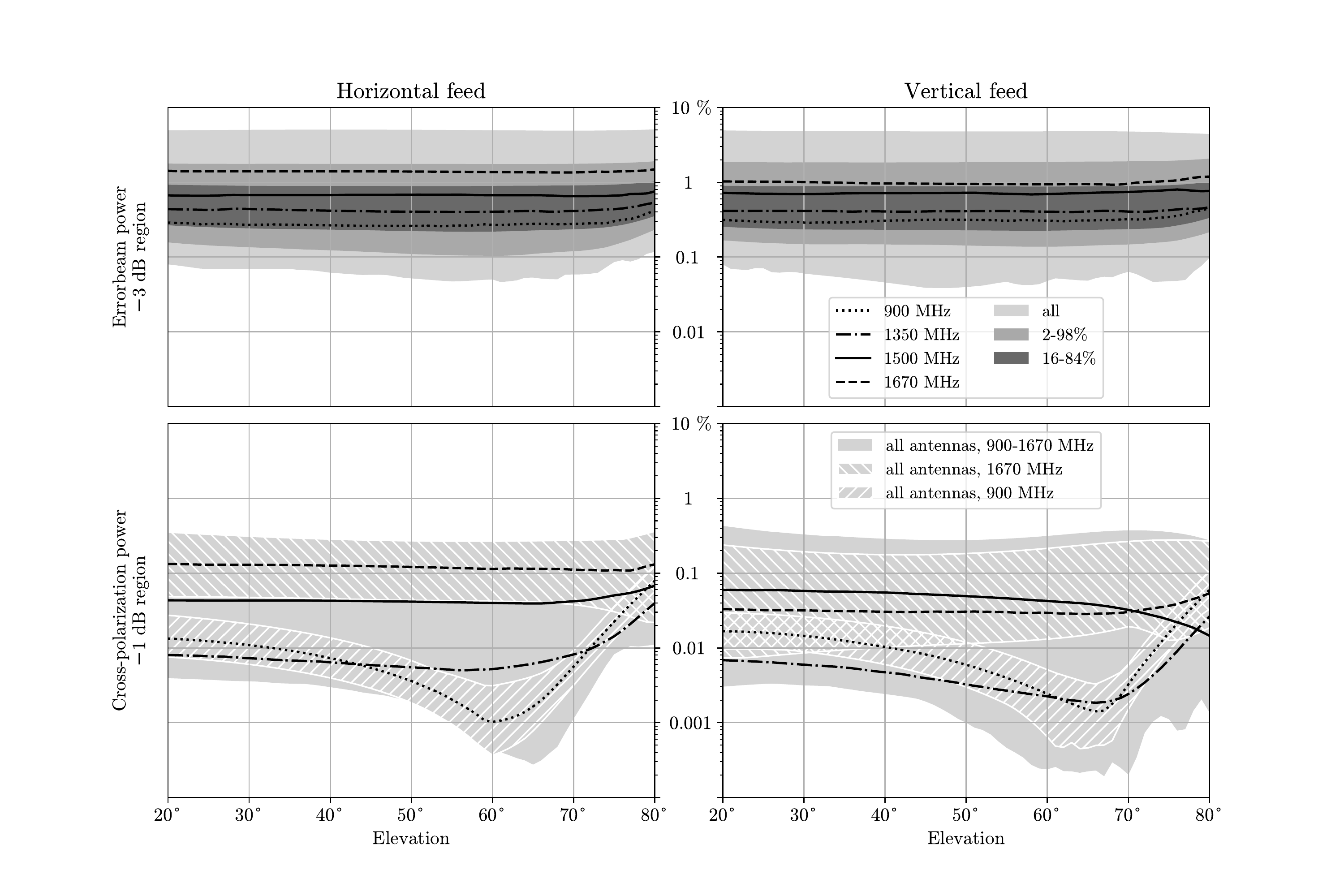}
\caption{Both errorbeam and cross-polarization level as a function of elevation show deterioration above $70^\circ$. At the lowest frequencies, cross-polarization patterns are more distinguished by elevation than to which antenna they belong.
\label{fig:error_elevation}}
\end{figure*}

Illustrated in the first row of Figure \ref{fig:error_elevation}, gravitational loading has a small effect on the copolarization power compared to antenna to antenna variations. Most commonly, gravitational loading introduces an error of 0.05\% over the $20^\circ$ to $70^\circ$ elevation range relative to the shape at $60^\circ$. At higher elevations an asymmetrical weight distribution appears to flex the feed support structure to the side. Although there is a notable effect on the patterns above 75$^\circ$, this effect still remains smaller than that of receiver to receiver differences. In contrast to the copolarization case, as seen in the second row of this figure, gravitational loading effects at low frequencies (indicated for 900 MHz) is much larger than antenna to antenna variations, for the cross-polarization patterns. At high frequencies (indicated for 1670 MHz) the gravitational loading effect is hardly noticeable against antenna to antenna variations. 

\begin{figure}[t!]
\centering
\includegraphics[width=\linewidth,trim=0.5cm 0.2cm 1.5cm 1.25cm, clip]{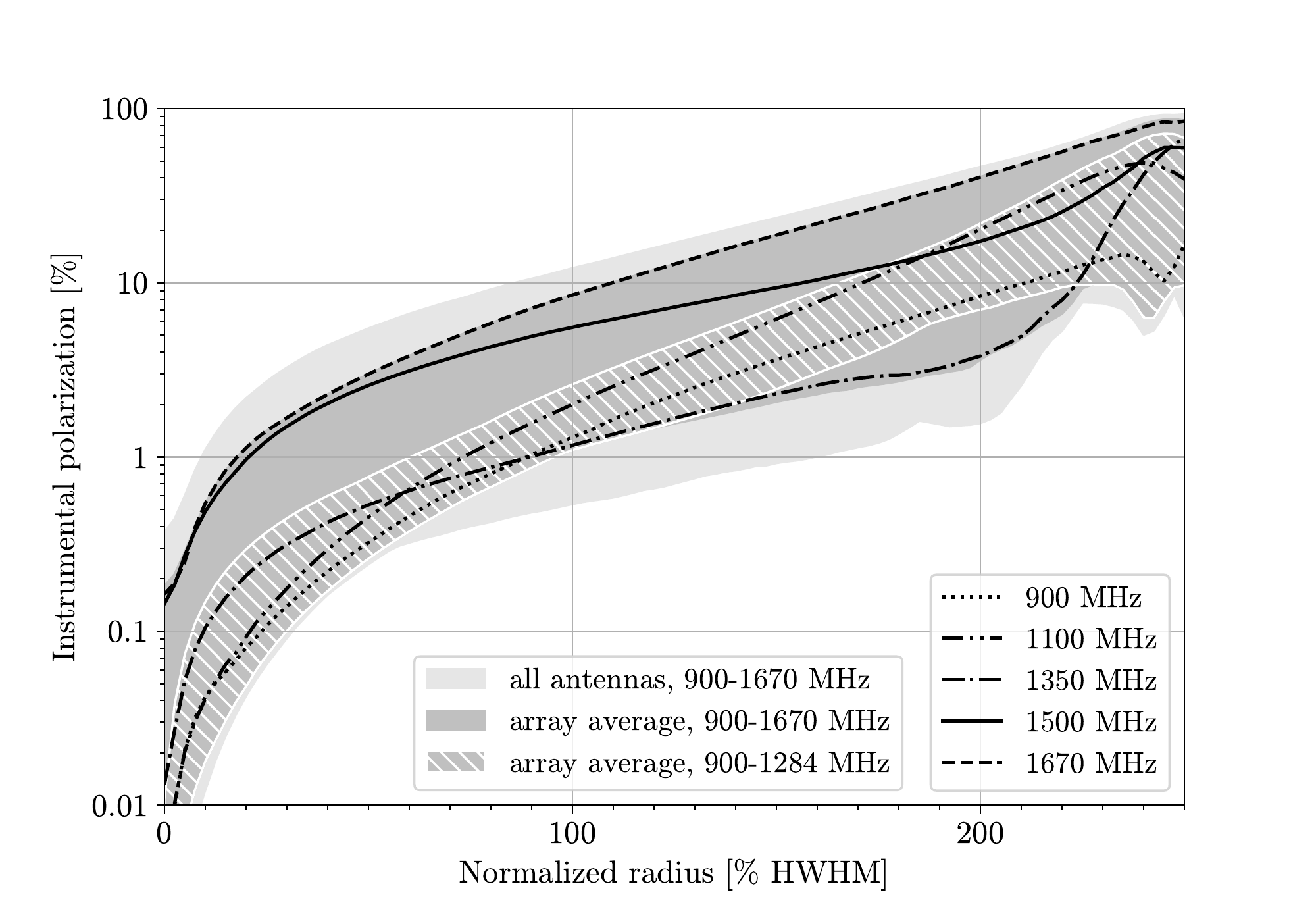}
\caption{The median (taken tangentially) of the instrumental polarization as a function of normalized radius for a number of frequencies. \label{fig:perc_pol}}
\end{figure}

The first two rows in Table \ref{tab:quv_errorbeam} show the maximum power levels of the Stokes Q, U, V response to an unpolarized source within the half power region of the Stokes I beam for all the antennas over the frequency ranges indicated. The polarization response above 1500 MHz is more than two times larger than in the lower half of the band. Errorbeam results in rows three and four show differences amongst antennas against the array average. While it is feasible to model the direction dependent polarization response using the array average in the lower half of the band to the 0.12-0.18\% level, on average, this can only be done to almost a factor of three times worse above 1500 MHz without taking into account receiver to receiver differences. Gravitational loading effects seems to be operating around similar power levels for these Stokes parameters as mentioned for Stokes I.

Figure \ref{fig:perc_pol} shows how the instrumental polarization deteriorates radially away from the beam center, particularly at higher frequencies, and towards the first null (which occurs radially at about 250\% of the HWHM). Within the half power region of the beam, the instrumental polarization is less than 2.5\% in the lower half of the band, but approaches 10\% in the upper half. Only in the lower half of the band is this performance comparable to that reported for the VLA by \cite{Cotton_1994}. At much higher operating frequencies and narrower fractional bandwidth, \cite{Hull_2020} measured up to 0.5\% instrumental polarization for the high precision ALMA submillimeter telescope near the half power level.
 
\begin{figure*}[ht!]
\centering
\includegraphics[width=0.78\linewidth,trim=2.5cm 1cm 2.8cm 1.8cm, clip]{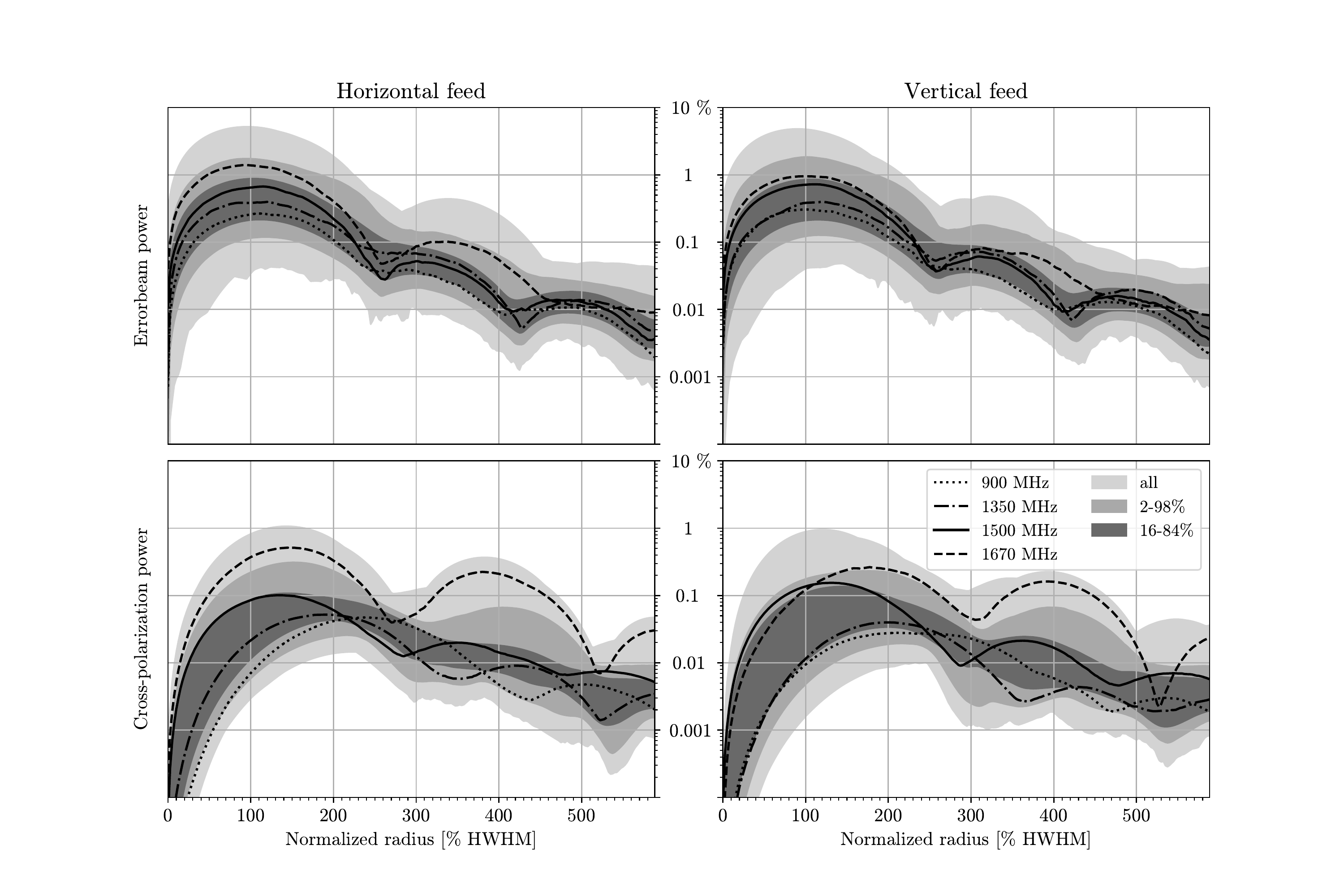}
\caption{The Errorbeam and cross-polarization power is illustrated as a function of radial distance from the beam center, collapsed tangentially, for all antennas and the restricted 900--1670 MHz frequency range. The styled lines indicate the median cases at specified frequencies. Note that the Stokes I beam incurs half power at the normalized radius value of 100\%.
\label{fig:error_radial}}
\end{figure*}

The distribution of error, and cross-polarization power along radial extent is shown in Figure \ref{fig:error_radial}. For various frequencies and antennas, the peak errorbeam is reached just before the half power point, while the peak cross-polarization power is reached just beyond the half power point. These effects are symptomatic of frequency dependent squint variations amongst individual antenna beams relative to the array average.

\begin{figure}[t!]
\centering
\includegraphics[width=\linewidth,trim=0.5cm 0.2cm 1.5cm 1.25cm, clip]{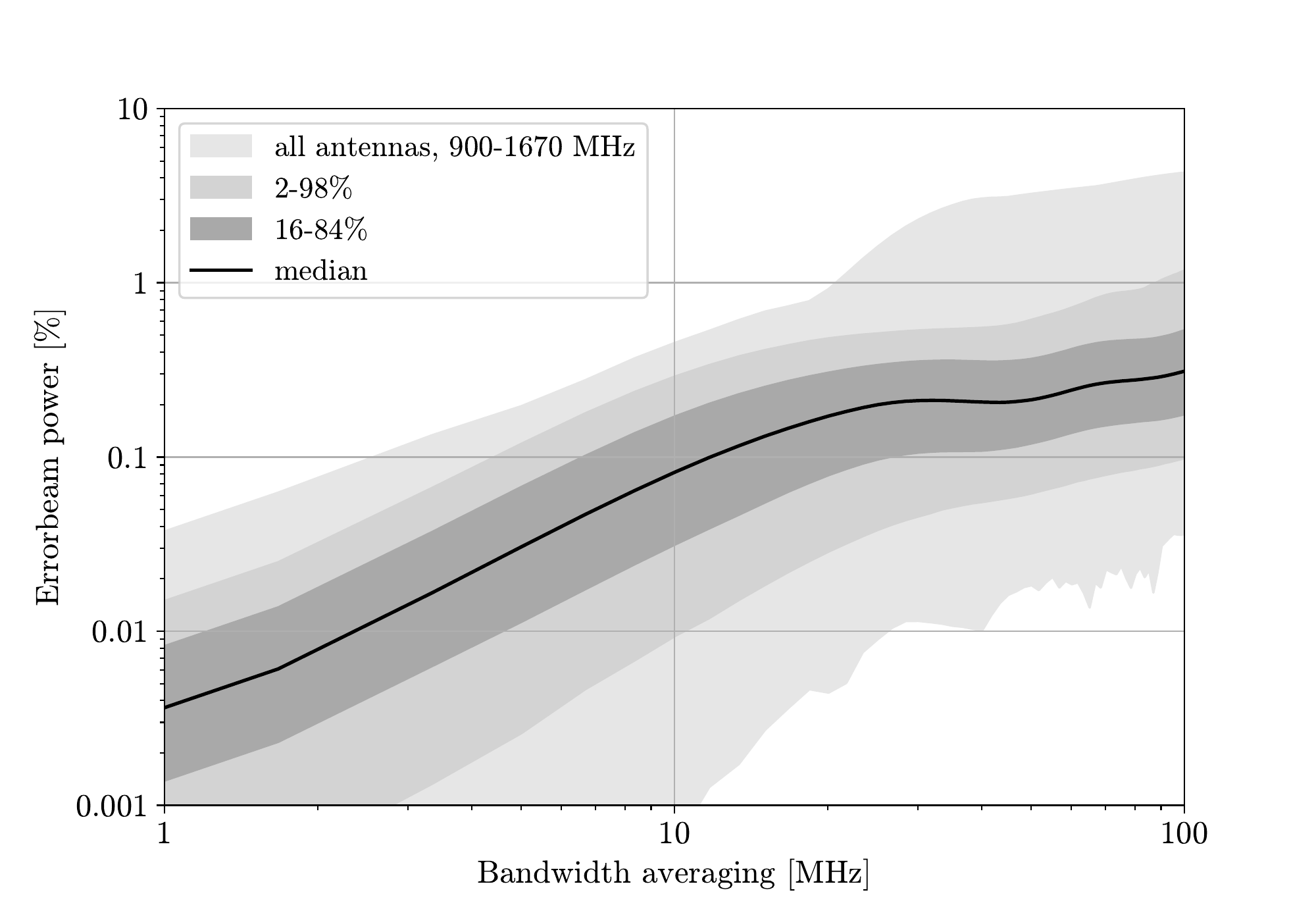}
\caption{The effect of bandwidth averaging gives a measure of the frequency quantization that can be budgeted for a given level of project accuracy. \label{fig:bandwidth_averaging}}
\end{figure}

\begin{figure*}[ht!]
\centering
\includegraphics[width=\linewidth,trim=2.5cm 0 1.7cm 0, clip]{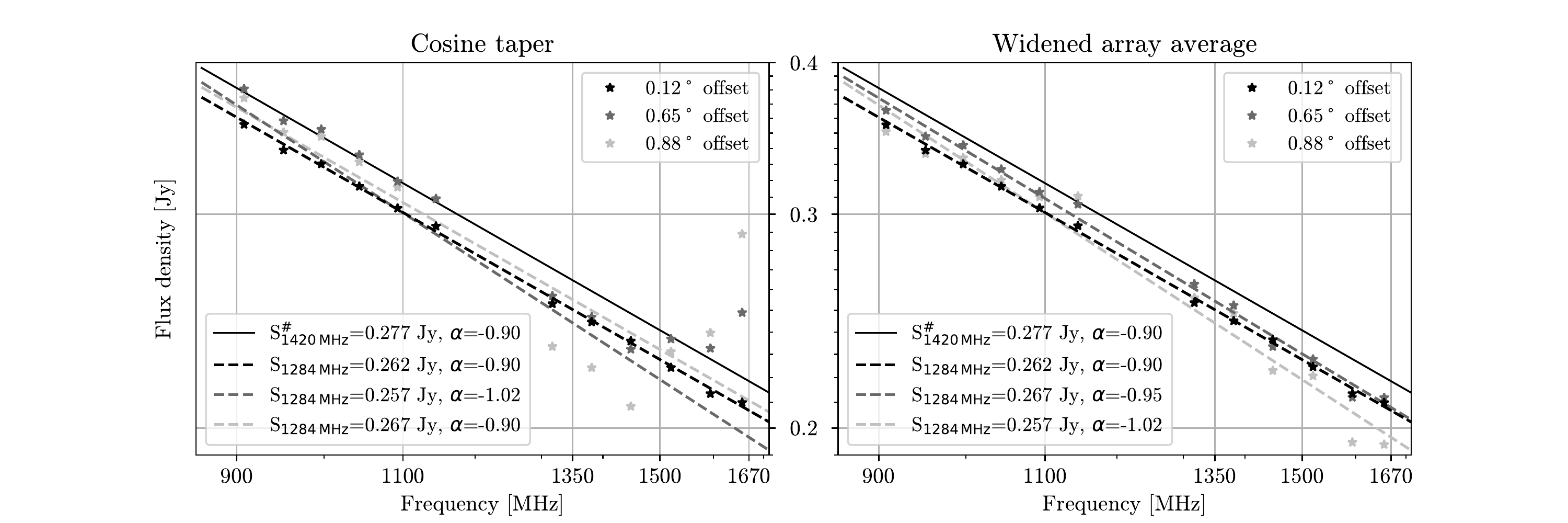}
\caption{Primary beam corrected log-log spectra for a test source J0038-7422 observed at 3 different offsets from the pointing, namely 0.12$^\circ$, 0.65$^\circ$ and 0.88$^\circ$. A power law spectrum is fitted to the points which is shown by dashed lines and parameters are given in legends; subband values are shown as $\star$. The solid line is for the literature reference marked with \# \citep{Volmer_2010}. Left panel is using the cosine beam and the right panel the widened array average beam. Noise error bars are smaller than the plotted symbols.}.
\label{fig:SIFig}
\end{figure*}

\subsection{Bandwidth averaging}
\label{section:bandwidth_averaging}
Interpolation of primary beam shapes across frequency is a task commonly performed by image correction algorithms, and is linked to a science project's  frequency resolution requirements and computer resource footprint. There are many ways in which a beam could be interpolated across frequency and the effect on accuracy for each method may differ significantly. Although it is recommended and would be more accurate to interpolate shape independently of its frequency dependent pointing offsets, doing so is unusual. 

To provide some guidance to the normal use case, i.e. without processing pointing behavior separately, the effect of bandwidth averaging on errorbeam is shown in Figure \ref{fig:bandwidth_averaging}. Here primary beams of all antennas at all frequencies in the band are compared to their average over a given bandwidth, using the errorbeam metric. For 10 MHz of bandwidth averaging, in most cases there is a maximum of only 0.08\% power introduced into the Stokes I beam. However, in the worst case for some antennas at some frequencies the maximum error could slightly exceed 0.4\%.

\subsection{HV phase residuals}

Over a timescale of 2 to 6 months it is very likely that an event happened where the digitizers of all antennas were power cycled. On shorter timescales of days to a month, digitizers might be reset during maintenance for some individual antennas. Whenever a digitizer starts up, a random delay of between $-$4 and 4 samples may occur in the alignment of one polarization signal relative to the other polarization of the same antenna (and other antennas). This is expected behavior based on how the phase lock on the maser timing signal is implemented in hardware and is generally inconsequential because user level calibration methods should eliminate these. However, polarization calibration that constrains the absolute position angle of a source is expected to yield multiples of 90$^\circ$ jumps in the leakage phase solutions per affected antenna from one dataset to another as a result of this effect, possibly causing some confusion.

In addition, there is a receiver specific frequency dependent residual in the phase alignment between polarizations. Its effect on polarization calibration is still under study.

\subsection{Geometric delay correction}

Prior to August 2019 and given sufficient signal to noise, visibility data captured by MeerKAT exhibit a distinctive sawtooth pattern in phase with a period of 5 seconds. This happens because geometric delay corrections (needed for ephemeris tracking) are interpolated and applied in hardware over this time period, and have a quantized delay rate resolution. Initially the phase of the visibilities could drift by up to about 6$^\circ$ within 5 seconds, depending on the baseline. Despite that the biasing effect on phase per baseline is small and is hard to identify in long dump rate observations, the quantization resolution has since been increased by 16 bits. A future release of the telescope's data access layer \texttt{katdal}\footnote{\href{https://github.com/ska-sa/katdal}{https://github.com/ska-sa/katdal}} intends to alleviate this issue retrospectively. 

On a related note, it may be worth cautioning that the standard system delay calibration performed when the array is prepared for an observation occasionally fails to determine delay offsets accurately for some antennas. It also happens (more rarely) that operators neglect to address this condition or do not mark such antennas as faulty. While the holography reductions check for and eliminate affected data, this could be a cause of degraded performance in other use cases.

\subsection{Spectral Index Example}
One of the cases where having an accurate beam shape is most critical is that of estimating spectral indices from primary beam corrected subband images. Since this involves data across the entire bandwidth, variations in antenna gain must be accurately and consistently applied. An example case comparing the cosine taper and a \em widened \em array averaged beam is shown in Figure \ref{fig:SIFig}.

This figure shows the same source, J0038-7422 \citep{Volmer_2010}, observed in extended ($\sim$8 hours) \em L-\em band observations at three different offsets from the nominal pointing. The offsets cover the range from inside the half power of the beam to well outside it. Subband flux densities of J0038-7422 are primary beam corrected and have a power law spectrum fitted (weighted by the inverse variance of the subband RMS). The left panel shows the result of using the simple cosine beam and the right panel uses an array average beam including pointing offsets. For the  0.12$^\circ$ offset (black stars), the two beam shapes give nearly identical results. At an offset of 0.65$^\circ$ (dark gray stars), the results are still quite similar however the spectrum corrected using the array average beam is closer to the 0.12$^\circ$ offset spectra than the one corrected using the cosine beam. At an offset of 0.88$^\circ$ (light gray stars) the cosine beam corrected spectrum is very wrong, rising sharply at higher frequencies, even though the fitted spectrum does not clearly indicate this. The spectrum corrected using the array average beam is a bit steeper than the 0.12$^\circ$ spectra and with considerably more scatter but is much closer to correct than the cosine corrected spectrum. 

The SNR of these data are all high enough that the deviations from a power law are due to inaccuracies in the assumed beam shapes. The array average beam shape is clearly preferred away from the center of the beam.  A simple power law fit to a grossly distorted spectrum may give misleading results. The cosine beam used does not include the effects of pointing errors which cause the array beam to broaden at higher frequencies where the pointing errors are a larger fraction of the beam size. The underestimate of the beam size inherent in this cosine beam causes the over corrections at high frequencies far from the nominal pointing.

\subsection{A perspective on use cases affecting errorbeam}

Table \ref{tab:snr_errorbeam_effects} summarizes various effects on the Stokes I beam accuracy, to express their relative magnitude and hence importance. Although most of these items have already been discussed earlier in this section, details in the statistics provided are expanded, and the results are collated for a more meaningful comparison. Reference is made to values quoted from Table \ref{tab:snr_errorbeam_effects} throughout this subsection to abet interpretation when read in a side by side fashion. The intent is to assist telescope users in estimating which effects need to be taken into account for a given level of beam accuracy that is appropriate to their project. Implications of the required primary beam accuracy may be far reaching on computing resources. 

Measurement accuracy statistics introduced in Section \ref{section:measurement_accuracy} are summarized in the first set of rows. In results of a typical single holography observation, corresponding to 30 minutes of observation time, most commonly there exists a maximum measurement error of about 0.29\% in power in the beam shape that is largely due to the signal to noise ratio of the measurement. If beam patterns for a single observation are averaged together into a half array average, the measurement error of this estimate of the array average is about 0.14\%. When combining results from multiple observations, the net per-antenna beam shape estimate is believed to be accurate to 0.05\% while the net array average is accurate to 0.02\%. These values give a sense to what degree of accuracy the beams are known currently.

Results quoted in the remaining rows of Table \ref{tab:snr_errorbeam_effects} are derived from comparisons against measured primary beam patterns for all antennas, or the array average (when thus indicated by $\dagger$), at an elevation of 60$^\circ$ and ambient temperature of 15$^\circ$C, over the 900 -- 1670 MHz frequency range. Antenna pointing errors are removed except in the two cases that deals with antenna pointing accuracy, while frequency dependent squint errors are retained.

Although higher order mode electromagnetic (EM) primary beam simulations from \cite{deVilliers_2021} are more exemplary than fundamental mode simulations, they too cannot compete with measurement-based modeling alternatives. This is because errorbeam results are sensitive to the actualized frequency dependent squint which cannot be predicted by simulations conclusively because of simulation approximations and manufacturing tolerances. The errorbeam valuations of EM simulations against all antennas in Table \ref{tab:snr_errorbeam_effects} are done only over 5 channels (at 900, 1100, 1350, 1500 and 1670 MHz), but these make up a representative sample. Note that the performance of these simulations are comparable to that of measurement-based modeling methods above 1500 MHz, discussed next.

Basic modeling of the beams can be done using a simple mathematical formula as in Section \ref{section:cosine_taper_beam_model}. If a cosine taper function, with frequency dependent pointing and beam width derived from the array average, is used to approximate all beam patterns, then errors of 0.87\% can be expected above 1500 MHz, while only 0.37\% in the lower half of the band, or 0.44\% on average. If instead frequency dependent pointing is tailored separately for each antenna from per-receiver squint measurements, then the error across the band reduces to only 0.25\%.

Errorbeam results improves by between 10 and 50\% if the array average pattern, instead of the cosine taper function, is used to approximate beam patterns for all antennas. By using the array average shape, but additionally tailoring the frequency dependent pointing separately for each antenna, then the error across the band reduces to 0.17\% on average. 

Beyond this level of modeling accuracy there is a small margin of benefit using individualized antenna beam shapes before elevation dependent effects demand consideration at 0.05\% power on average (Section \ref{section:power_levels_of_errorbeam}). In terms of cross-polarization, there is more variability due to elevation than per-antenna differences in the lower half of the band. In the upper half of the band, beam patterns are associated more representatively by which receiver is installed on the antenna regardless of the antenna. Using beams at this level of accuracy necessitates more complex bookkeeping and it may be computationally demanding to apply in imaging software.

By far the most striking outlier effect encountered is due to feed indexer slippage that happened to 4 antennas to a significant degree discussed towards the end of Section \ref{section:the_measured_copolarization_beams}. Slippage causes devastating pointing errors that reoccur despite being compensated by routine system pointing calibration. Apart from resulting pointing errors, the loss in collimation also affects the beam shape with coma distortion, which in the tabulated case reads a 1.83\% error due to the gradual change in shape alone for M054 over 2$\frac{1}{4}$ years. In June 2021, a sudden further slippage from 3.2$^\circ$ to 4.4$^\circ$ relative to its position at assembly, sparked a thorough investigation and subsequent mechanical fix by early July. In some applications it may be more sensible to reject data from antennas that differ by much from the array average, instead of individualizing beam patterns.

As indicated in Section \ref{section:pointingerror_errorbeam}, for typical observations the system is unable to point antennas with an accuracy closer than $\sigma=0.64$ arcminutes to an intended target. This has the consequence that telescope users cannot align beam patterns accurately, resulting in 1.16\% errors in the applied beam patterns if no additional provisions are made. This hurdle may be the stumbling block for most telescope users, making it pointless in this case to use high accuracy beams.

Those users that perform post-imaging primary beam corrections, should be aware that there remains a larger than expected $\sigma=0.25$ arcminute inaccuracy to the array average pointing. The assumption that the array average primary beam has no pointing error would typically introduce an error of up to 0.39\% in the assumed beam power near the half power point. Refer to Figures \ref{fig:antpointing_history} and \ref{fig:pointing_errorbeam}.

The compounding of lesser effects such as bandwidth averaging from Section \ref{section:bandwidth_averaging}, gravitational loading and night time ambient temperature could easily induce deviations in the beam shape above the 0.1\% power level. Because the receiver to receiver deviations from the array average in the lower half of the band is also at that level for Stokes Q, U, V beams according to Table \ref{tab:quv_errorbeam}, polarimetry studies will not benefit from per-antenna patterns in the lower half of the band, unless these lesser affects are also included. However, above 1500 MHz, it is necessary to regard receiver dependence in order to achieve an accuracy below the 0.3\% power level for Stokes Q, U, V beam models. 

It is advised that investigators first achieve satisfactory results in the lower half of the band, where possible, before attempting the upper half which is more challenging due to effects of the activation of higher order waveguide modes.

\section{Conclusion}

A great level of detail and accuracy is provided on the measured characteristics of the MeerKAT primary beam shape. These are presented as figures for reference. Perhaps the polarization and frequency dependent pointing behavior is the most interesting feature, which is due to the activation of higher order waveguide modes in the OMT. Receiver to receiver variations in the beam shape are significantly boosted in the upper half of the band by this unexpected effect. However this is currently not the biggest factor that limits accuracy for most users.

The most important consideration in utilizing MeerKAT primary beams effectively is the manner in which pointing errors are dealt with. If the model beam is not aligned accurately to where the antenna is truly pointing, then it introduces an error in the presumed beam shape that could quickly overwhelm any other effect on the beam that may occur during normal telescope operations. Given the pointing calibration accuracy that MeerKAT is currently capable of, the assumption that the beams are mechanically aligned towards the intended pointing center of a target field, is roughly a two times poorer assumption (in terms of maximum error power introduced into the beam) than using a well-oriented cosine taper to approximate the beam shape. It is crucial that MeerKAT's system pointing be improved or augmented in order to make effective use of beam patterns that exceed an accuracy of 1\% in power.

There are two distinct categories of use for the primary beams in imaging. In the case that is by far more common, a \em widened \em array average beam pattern is required to correct attenuation in final images so that more accurate flux densities and spectral indices can be derived across the field of view. Alternatively, in the case to improve image quality instead, specialist imaging deconvolution algorithms that iterate between the modeled sky and visibilities using accurate beam models, can estimate pointing errors at an enormous processing expense. The accuracy to which the beam pointing can be estimated depends on source locations in the field, uncertainties of source fluxes and the beam model; still that solution may be entangled with other calibration parameters. 

Pipeline imaging and common telescope usage rely more heavily on the system pointing accuracy because the computational overhead of the alternative is largely prohibitive. It is up to the telescope science user to assess the available resources and intricacies of their software to determine what effects are feasible to be taken into account and to estimate what level of accuracy can be achieved. In some cases the pointing error will remain the limiting factor rather than the accuracy of the beam model used.

At a level secondary to ordinary antenna pointing, the frequency dependent nature of the pointing errors or simplifying assumptions regarding its variability may limit the accuracy to which telescope users can utilize accurate beam patterns. In the lower half of the band, antenna to antenna variations in beam pattern shape are about three times smaller than above 1500 MHz. For the copolarization beams this is largely accounted for by frequency depending pointing differences linked to the receivers installed, rather than shape differences due to collimation. In most cases it is possible to represent antenna beam patterns to less than 0.2\% maximum error in power by recentering the array average beam shape using frequency dependent pointing details peculiar to specific receivers installed. Without tailoring frequency dependent pointing errors per antenna, an array average beam shape would mostly be 0.3\% in error, but around 0.8\% above 1500 MHz.

At a third level of accuracy beam shapes are affected by gravitational loading due to the antenna elevation. Cross-polarization beam shapes are more affected by gravitational loading than antenna to antenna differences in the lower half of the band where only the fundamental waveguide mode propagates. Computational provisions need to be expanded to use different beam shapes based on antenna, the elevation, and depending on the requirements also temperature, for the highest accuracy beam corrections. At this level of accuracy dated beam patterns are needed because receivers are exchanged, and feed indexer encoder slippage and collimation adjustments occur over a period of time. Alternatively, in some applications, data from antennas that are not well represented by the array average could be eliminated.

\newpage
\begin{acknowledgments}
This study is enabled by a large number of people at SARAO involved in the design, build, maintenance and running of the telescope, notably the telescope operators and astronomers on duty that facilitated the scheduling and execution of most of the observations at odd hours. Special thanks to Lance Williams for streamlining the use of the antenna proxy when buffering requested coordinates for arbitrary scan patterns. Henno Kriel for confirming behavioral details of the digitizer. Ludwig Schwardt for blackbelt help in writing Python modules, and timely fixes to platform software problems that arise. Adriaan Peens-Hough for providing system specification details and technical user feedback. Pieter P. A. Kotz\'e and Benjamin Lunsky for getting new unqualified antennas operational. Matthys Maree, Gerhard Botha, Dalton Taylor and Raymond Malgas for on site inspections and adjustments. 

This paper was subject to an internal review. Thanks to Fernando Camilo, Ludwig Schwardt, Thomas Abbott, Justin Jonas, Oleg Smirnov and Sharmila Goedhart for consideration and feedback.

The MeerKAT telescope is operated by the South African Radio Astronomy Observatory, which is a facility of the National Research Foundation, an agency of the Department of Science and Innovation. The National Radio Astronomy Observatory is a facility of the National Science Foundation, operated under a cooperative agreement by Associated Universities, Inc.
\end{acknowledgments}

\bibliography{mkat_beam}{}

\begin{thebibliography}{}
\expandafter\ifx\csname natexlab\endcsname\relax\def\natexlab#1{#1}\fi
\providecommand{\url}[1]{\href{#1}{#1}}
\providecommand{\dodoi}[1]{doi:~\href{http://doi.org/#1}{\nolinkurl{#1}}}
\providecommand{\doeprint}[1]{\href{http://ascl.net/#1}{\nolinkurl{http://ascl.net/#1}}}
\providecommand{\doarXiv}[1]{\href{https://arxiv.org/abs/#1}{\nolinkurl{https://arxiv.org/abs/#1}}}

\bibitem[{Asad {et~al.}(2021)Asad, Girard, de Villiers, Ansah-Narh, Iheanetu,
  Smirnov, Santos, Lehmensiek, Jonas, de Villiers, Thorat, Hugo, Makhathini,
  Jozsa, \& Sirothia}]{Asad_2019}
Asad, K. M.~B., Girard, J.~N., de Villiers, M., {et~al.} 2021, MNRAS, 502,
  2970, \dodoi{10.1093/mnras/stab104}

\bibitem[{{Condon} \& {Ransom}(2016)}]{Condon_2016}
{Condon}, J.~J., \& {Ransom}, S.~M. 2016, {Essential Radio Astronomy}
  (Princeton, NJ: Princeton University Press)

\bibitem[{{Cotton}(1994)}]{Cotton_1994}
{Cotton}, W.~D. 1994, AIPS Memo, 86

\bibitem[{Cotton \& Mauch(2021)}]{Cotton_2021}
Cotton, W.~D., \& Mauch, T. 2021, PASP, 133, 104502,
  \dodoi{10.1088/1538-3873/ac2351}

\bibitem[{{de Villiers} {et~al.}(2021){de Villiers}, {Venter}, \&
  {Peens-Hough}}]{deVilliers_2021}
{de Villiers}, M.~S., {Venter}, M., \& {Peens-Hough}, A. 2021, ITAP, 69, 6333,
  \dodoi{10.1109/TAP.2021.3070110}

\bibitem[{Harp {et~al.}(2011)Harp, Ackermann, Nadler, Blair, Davis, Wright,
  Forster, DeBoer, Welch, Atkinson, Backer, Backus, Barott, Bauermeister,
  Blitz, Bock, Bower, Bradford, Cheng, Croft, Dexter, Dreher, Engargiola,
  Fields, Heiles, Helfer, Jordan, Jorgensen, Kilsdonk, Gutierrez-Kraybill,
  Keating, Law, Lugten, MacMahon, McMahon, Milgrome, Siemion, Smolek, Thornton,
  Pierson, Randall, Ross, Shostak, Tarter, Urry, Werthimer, Williams, \&
  Whysong}]{Harp_2011}
Harp, G.~R., Ackermann, R.~F., Nadler, Z.~J., {et~al.} 2011, ITAP, 59, 2004,
  \dodoi{10.1109/TAP.2011.2122214}

\bibitem[{Hull {et~al.}(2020)Hull, Cortes, Gouellec, Girart, Nagai, Nakanishi,
  Kameno, Fomalont, Brogan, Moellenbrock, \& et~al.}]{Hull_2020}
Hull, C. L.~H., Cortes, P.~C., Gouellec, V. J. M.~L., {et~al.} 2020, PASP, 132,
  094501, \dodoi{10.1088/1538-3873/ab99cd}

\bibitem[{Iheanetu {et~al.}(2019)Iheanetu, Girard, Smirnov, Asad, de Villiers,
  Thorat, Makhathini, \& Perley}]{Iheanetu_2019}
Iheanetu, K., Girard, J.~N., Smirnov, O., {et~al.} 2019, MNRAS, 485, 4107,
  \dodoi{10.1093/mnras/stz702}

\bibitem[{Mauch {et~al.}(2020)Mauch, Cotton, Condon, Matthews, Abbott, Adam,
  Aldera, Asad, Bauermeister, Bennett, \& et~al.}]{Mauch_2020}
Mauch, T., Cotton, W.~D., Condon, J.~J., {et~al.} 2020, ApJ, 888, 61,
  \dodoi{10.3847/1538-4357/ab5d2d}

\bibitem[{{Perley}(2016)}]{Perley_2016}
{Perley}, R. 2016, EVLA memo, 195

\bibitem[{{Perley}(2021)}]{Perley_2021}
---. 2021, EVLA memo, 212

\bibitem[{{Popping} \& {Braun}(2008)}]{Popping_2008}
{Popping}, A., \& {Braun}, R. 2008, A\&A, 479, 903,
  \dodoi{10.1051/0004-6361:20079122}

\bibitem[{{Smirnov}(2011)}]{Smirnov_2011}
{Smirnov}, O.~M. 2011, A\&A, 527, A106, \dodoi{10.1051/0004-6361/201016082}

\bibitem[{Tingay {et~al.}(2003)Tingay, Jauncey, King, Tzioumis, Lovell, \&
  Edwards}]{Tingay_2003}
Tingay, S.~J., Jauncey, D.~L., King, E.~A., {et~al.} 2003, PASJ, 55, 351,
  \dodoi{10.1093/pasj/55.2.351}

\bibitem[{{Vollmer} {et~al.}(2010){Vollmer}, {Gassmann}, {Derri\`ere}, {Boch},
  {Louys}, {Bonnarel}, {Dubois}, {Genova}, \& {Ochsenbein}}]{Volmer_2010}
{Vollmer}, B., {Gassmann}, B., {Derri\`ere}, S., {et~al.} 2010, A\&A, 511, A53,
  \dodoi{10.1051/0004-6361/200913460}

\end{thebibliography}
\bibliographystyle{aasjournal}

\begin{deluxetable*}{ccccc}
\tablenum{2}
\tablecaption{A perspective on Stokes I Errorbeam power levels\label{tab:snr_errorbeam_effects}}
\tablewidth{0pt}
\tablehead{
 &\colhead{Description} & \multicolumn3c{Errorbeam power [\%]} \\
 &\colhead{} & \colhead{median} & \colhead{16-84$^\mathrm{th}$ percentile} & \colhead{2-98$^\mathrm{th}$ percentile}
}
\startdata
\multirow{4}{*}{\rotatebox[origin=c]{90}{\parbox{2.2cm}{\centering Measurement\\accuracy}}}&Antenna, individual, SNR 60 & 0.29 & 0.17-0.57& 0.11-2.59\\
&Half array, indiscriminate, SNR 60 & 0.14  &0.08-0.26& 0.05-1.26\\
&Half array, consolidated, SNR 60 & 0.05  &0.02-0.10& 0.01-0.32\\
&Antenna, composite & 0.05 &0.03-0.09&0.02-0.14\\
&Array average, composite & 0.02  &0.01-0.04&0.01-0.05\\
\hline
\multirow{8}{*}{\rotatebox[origin=c]{90}{\parbox{4.2cm}{\centering Modeling simplifications}}}
&EM simulations, fundamental mode, 5 channels only&0.88&0.55-2.89&0.41-4.39\\
&EM simulations, high order, 5 channels only&0.75&0.53-1.57&0.43-2.60\\
\cline{2-5}
&Cosine taper, array average squint, $>$1500 MHz &0.87 &0.55-1.39&0.36-2.32\\
&Cosine taper, array average squint &0.44  &0.30-0.86&0.23-1.67\\
&Cosine taper, array average squint, $<$1284 MHz &0.37  &0.28-0.50&0.23-0.94\\
&Cosine taper, per-antenna squint & 0.25 &0.21-0.37&0.18-0.84\\
\cline{2-5}
&Array average, $>$1500 MHz &0.78&0.48-1.35&0.28-2.20\\
&Array average &0.34&0.19-0.77&0.11-1.62\\
&Array average, $<$1284 MHz &0.25&0.16-0.40&0.09-0.97\\
&Array average shape, per-antenna squint& 0.17 & 0.10-0.33 & 0.06-1.14\\
\hline
\multirow{7}{*}{\rotatebox[origin=c]{90}{\parbox{1.5cm}{Use cases}}}&Feed indexer slippage, 3.2$^{\circ}$ $\dagger$& 1.83 & 1.34-2.18 &1.17-3.31\\
\cline{2-5}
&Antenna pointing accuracy, $\sigma = 0.64$ arcmin & 1.16  & 0.59-3.03 & 0.26-4.55\\
&Array pointing accuracy, $\sigma = 0.25$ arcmin $\dagger$& 0.39  & 0.19-0.70 & 0.07-1.06\\
\cline{2-5}
&Bandwidth averaging, 10 MHz & 0.08 &0.03-0.17&0.01-0.29\\
\cline{2-5}
&Gravitational load, $70^{\circ}$ -- $80^{\circ}$ elevation $\dagger$& 0.11 &0.07-0.17&0.05-0.23\\
&Gravitational load, $20^{\circ}$ -- $70^{\circ}$ elevation $\dagger$& 0.05 &0.02-0.09&0.00-0.14\\
\cline{2-5}
&Nocturnal thermal contraction, $\Delta10 ^{\circ}$C $\dagger$& 0.03 &0.02-0.03&0.02-0.04\\
\cline{2-5}
&Feed indexer consistency, $\sigma = 0.62$ millidegrees & 0.01 & 0.00-0.02& 0.00-0.04\\
\enddata
\tablecomments{Statistics are evaluated over Stokes I beams for all antennas, except for items marked with $\dagger$ where only the array average is considered. Antenna pointing errors are removed in all cases except for the two items that quote errors due to pointing accuracy. Performance outside the 900 -- 1670 MHz system specification range deteriorates and is excluded.}
\end{deluxetable*}

\begin{deluxetable*}{cccccccc}
\tablenum{3}
\tablecaption{Variability in Stokes QUV\label{tab:quv_errorbeam}}
\tablewidth{0pt}
\tablehead{
 \colhead{}& \colhead{Description} & \multicolumn2c{Stokes Q [\%]}& \multicolumn2c{Stokes U [\%]} & \multicolumn2c{Stokes V [\%]}\\
 \colhead{}&\colhead{} & \colhead{median} & \colhead{16-84$^\mathrm{th}$ percentile} & \colhead{median} & \colhead{16-84$^\mathrm{th}$ percentile} & \colhead{median} & \colhead{16-84$^\mathrm{th}$ percentile}
}
\startdata
\multirow{2}{*}{\rotatebox[origin=c]{90}{\parbox{0.8cm}{}}}&Maximum value, $>$1500 MHz & 2.21 & 1.88-3.29  & 2.79 & 2.32-3.71 & 2.23 & 1.93-2.63  \\
&Maximum value, $<$1284 MHz & 1.09 & 0.84-1.38  & 1.05 & 0.77-1.29 & 0.33 & 0.19-0.49  \\
\hline
\multirow{4}{*}{\rotatebox[origin=c]{90}{\parbox{1.5cm}{Errorbeam}}}&Array average, $>$1500 MHz & 0.32 & 0.17-0.69  & 0.43 & 0.22-0.74 & 0.39 & 0.20-0.70  \\
&Array average, $<$1284 MHz & 0.12 & 0.07-0.21  & 0.14 & 0.08-0.24 & 0.18 & 0.10-0.37  \\
&Elevation, $70^\circ$ -- $80^\circ$ $\dagger$ & 0.03 & 0.02-0.05  & 0.07 & 0.04-0.13 & 0.07 & 0.04-0.14  \\
&Elevation, $20^\circ$ -- $70^\circ$ $\dagger$& 0.02 & 0.01-0.04  & 0.04 & 0.01-0.08 & 0.03 & 0.01-0.05  \\
\enddata
\tablecomments{All values are expressed as a percentage power relative to the Stokes I maximum. Statistics are evaluated for all antennas, except for items marked with $\dagger$ where only the array average is considered. Performance evaluation is restricted to within the 900 -- 1670 MHz range.}
\end{deluxetable*}

\end{document}